\documentclass[%
 reprint,
 superscriptaddress,
 nofootinbib,
 amsmath,amssymb,
 aps,
 prd,
 eqsecnum
]{revtex4-2}

\usepackage[T1]{fontenc}
\usepackage{graphicx}
\usepackage{dcolumn}
\usepackage{color}
\usepackage{amsfonts}
\usepackage{mathrsfs}
\usepackage{bm}
\usepackage{bbm}
\usepackage[caption=false]{subfig}
\usepackage{xspace}
\usepackage{hyperref}
\usepackage{cleveref}

\DeclareMathOperator{\arccosh}{arccosh}
\DeclareMathOperator{\arcsinh}{arcsinh}

\def\dd{\text{d}}
\def\pb{\bar{p}}
\def\muIR{\mu_{\text{IR}}}
\def\euler{\gamma_{\text{E}}}
\def\ie{i\epsilon}
\def\y{y}
\def\arccoshy{\arccosh\y} 
\def\arcsinhw1q2{\arcsinh\frac{-w_1}{\sqrt{q_2^2}}}
\def\MKMOC{\mathscr{M}}
\def\WEFT{\mathcal{W}}

\def\WEFTb{W_{\text{EFT}}}

\def\opT{\mathbb{T}}
\def\opO{\mathbb{O}}
\def\opa{\hat{a}}
\def\hdelta{\hat{\delta}}
\def\FT{\text{FT}_{b}}
\def\Smatrix{\mathbb{S}}
\newcommand{\com}{$\text{c.m.}$\xspace}
\newcommand{\incom}{incoming \com}

\newcommand{\barcom}{barred \com}

\def\ME{\mathsf{M}}

\crefname{equation}{Eq.}{Eqs.}
\crefname{section}{Sec.}{Secs.}
\crefname{figure}{Fig.}{Figs.}
\crefname{appendix}{App.}{Apps.}

\newcommand{\be}{\begin{equation}}
\newcommand{\ee}{\end{equation}}
\newcommand{\beq}{\begin{equation}}
\newcommand{\eeq}{\end{equation}}
\newcommand{\bea}{\begin{eqnarray}}
\newcommand{\eea}{\end{eqnarray}}
\def\nn{\nonumber}
\newcommand{\eq}{\eqref}

\newcommand{\om}{{\omega}}

\newcommand{\eps}{\epsilon}

\newcommand{\mb}{\bar m}
\newcommand{\pinf}{p_{\infty}}
\newcommand{\D}{\partial}
\newcommand{\n}{\mathbf{n}}
\newcommand{\p}{\mathbf{p}}

\usepackage{tikz}

\usetikzlibrary{calc}
\usetikzlibrary{decorations.markings}
\usetikzlibrary{decorations.pathmorphing}

\tikzset{
    graviton/.style={decorate,line width=0.15mm, 
	decoration={snake,amplitude=.6mm, segment length=2mm}
        },
    massive/.style={postaction={decorate},
		line width=0.75mm,
	},
    massive2/.style={postaction={decorate},
		line width=0.3mm,
	}
}

\tikzset{
    treeAmp/.pic={
        \pgfmathsetmacro{\r}{2};        
        \foreach \x in {-45,-135} {
            \draw [massive] (0,0) -- ++ (\x:\r);
        }
        \foreach \x in {45,135} {
            \draw [massive2] (0,0) -- ++ (\x:\r);
        }
        \draw [graviton] (0,0) -- (\r,0);
        \filldraw [fill=gray!50!white] (0,0) circle (0.5cm);
    },
    oneLoopAmp/.pic={
        \pgfmathsetmacro{\r}{2};        
        \foreach \x in {-45,-135} {
            \draw [massive] (0,0) -- ++ (\x:\r);
        }
        \foreach \x in {45,135} {
            \draw [massive2] (0,0) -- ++ (\x:\r);
        }
        \draw [graviton] (0,0) -- (\r,0);
        \filldraw [fill=gray!50!white] (0,0) circle (0.75cm);
        \filldraw [fill=white] (0,0) circle (0.35cm);
    },
    oneLoopCut/.pic={
        \pgfmathsetmacro{\h}{2.8}
        \pgfmathsetmacro{\w}{2}
        \draw [massive2] (-\w/2,0) -- (-\w,\h/2) (-\w/2,0) to[bend left=60] (\w/2,0) (\w/2,0) -- (\w,\h/2) ;
        \draw [massive] (-\w/2,0) -- (-\w,-\h/2) (-\w/2,0) to[bend right=60] (\w/2,0) (\w/2,0) -- (\w,-\h/2) ;
        \draw [graviton] (-\w/2,0) to[bend left=45] (0,\h/2);
        \filldraw [fill=gray!50!white,thick] (-\w/2,0) circle (0.35cm) (\w/2,0) circle (0.35cm);
        \draw [line width=3pt,draw=white] (0,\h/2+0.1) -- (0,-\h/2-0.1);
        \draw [dashed] (0,\h/2+0.1) -- (0,-\h/2-0.1);
    },
    discTree/.pic={
        \pgfmathsetmacro{\h}{1}
        \pgfmathsetmacro{\w}{2}
        \pgfmathsetmacro{\hb}{1}
        \draw [line width=3pt,draw=white] (0,\h+\hb+0.1) -- (0,-0.1);
        \draw [massive] (-\w,0) -- (\w,0);
        \draw [massive2] (-\w,\h) -- (\w,\h);
        \draw [graviton] (0,\h) -- (\w,\h+\hb);     
        \node at (-\w,0) [left=0pt] {$p_1$};
        \node at (-\w,\h) [left=0pt] {$p_2$};
        \node at (\w,0) [right=0pt] {$p_4$};
        \node at (\w,\h) [right=0pt] {$p_3$};
        \node at (\w,\h+\hb) [right=0pt] {$k$};
    },
    discOneLoopBG/.pic={
        \pgfmathsetmacro{\h}{1}
        \pgfmathsetmacro{\w}{2}
        \pgfmathsetmacro{\hb}{1}
        \draw [massive] (-\w,0) -- (\w,0);
        \draw [massive2] (-\w,\h) -- (\w,\h);
        \draw [graviton] (-\w/2,\h) -- (0,\h+\hb);
        \coordinate (a) at (-\w/2,\h/2);
        \draw [graviton] (a) to[in=-135,out=0] (\w/2,\h);
        \draw [line width=3pt,draw=white] (0,\h+\hb+0.1) -- (0,-0.1);
        \draw [dashed] (0,\h+\hb+0.1) -- (0,-0.1);
        \filldraw [fill=gray!50!white] (-\w/2,\h/2) ellipse (0.2 and \h/2+0.1);
        \node at (-\w,0) [left=0pt] {$p_1$};
        \node at (-\w,\h) [left=0pt] {$p_2$};
        \node at (\w,0) [right=0pt] {$p_4$};
        \node at (\w,\h) [right=0pt] {$p_3$};
        \node at (0,\h+\hb) [right=0pt] {$k$};
    },
    discOneLoopCut/.pic={
    \pgfmathsetmacro{\h}{1}
        \pgfmathsetmacro{\w}{2}
        \pgfmathsetmacro{\hb}{1}
        \draw [massive] (-\w,0) -- (\w,0);
        \draw [massive2] (-\w,\h) -- (\w,\h);
        \draw [graviton] (-\w/2,\h) -- (0,\h+\hb);
        \draw [line width=3pt,draw=white] (0,\h+\hb+0.1) -- (0,-0.1);
        \draw [dashed] (0,\h+\hb+0.1) -- (0,-0.1);
        \filldraw [fill=gray!50!white] (\w/2,\h/2) ellipse (0.2 and \h/2+0.1);
    },
    discOneLoopCutDS/.pic={
    \pgfmathsetmacro{\h}{1}
        \pgfmathsetmacro{\w}{2}
        \pgfmathsetmacro{\hb}{1}
        \draw [massive] (-\w,0) -- (\w,0);
        \draw [massive2] (-\w,\h) -- (\w,\h);
        \draw [graviton] (-\w/2,\h) -- (0,\h+\hb);
        \coordinate (a) at (\w/2,\h/2);
        \draw [graviton] (a) to[in=-45,out=180] (-\w/2,\h);
        \draw [line width=3pt,draw=white] (0,\h+\hb+0.1) -- (0,-0.1);
        \draw [dashed] (0,\h+\hb+0.1) -- (0,-0.1);
        \filldraw [fill=gray!50!white] (\w/2,\h/2) ellipse (0.2 and \h/2+0.1);
    },
    discOneLoopCutDS2/.pic={
    \pgfmathsetmacro{\h}{1}
        \pgfmathsetmacro{\w}{2}
        \pgfmathsetmacro{\hb}{1}
        \draw [massive] (-\w,0) -- (\w,0);
        \draw [massive2] (-\w,\h) -- (\w,\h);
        \draw [graviton] (-\w/2,\h) -- (0,\h+\hb);
        \coordinate (a) at (\w/2,\h/2);
        \draw [graviton] (a) to[in=45,out=180] (-\w/2,0);
        \draw [line width=3pt,draw=white] (0,\h+\hb+0.1) -- (0,-0.1);
        \draw [dashed] (0,\h+\hb+0.1) -- (0,-0.1);
        \filldraw [fill=gray!50!white] (\w/2,\h/2) ellipse (0.2 and \h/2+0.1);
    }
}

\begin{document}

\title{Gravitational Waveform: A Tale of Two Formalisms}%

\author{Donato~Bini}
\affiliation{
    Istituto per le Applicazioni del Calcolo M. Picone, CNR, I-00185 Rome, Italy
}
\affiliation{
    INFN, Sezione di Roma Tre, I-00146 Rome, Italy
}
\author{Thibault~Damour}
\affiliation{
    Institut des Hautes Etudes Scientifiques, 
    91440 Bures-sur-Yvette, France
}
\author{Stefano~De~Angelis}
\affiliation{
    Institut de Physique Th\'eorique, CEA, CNRS, Universit\'e Paris-Saclay,
    F–91191 Gif-sur-Yvette cedex, France
}
\author{Andrea~Geralico}
\affiliation{
    Istituto per le Applicazioni del Calcolo M. Picone, CNR, I-00185 Rome, Italy
}
\author{Aidan~Herderschee}
\affiliation{
    Institute for Advanced Study, 
    Princeton, NJ 08540, USA
}
\author{Radu~Roiban}
\affiliation{
    Institute for Gravitation and the Cosmos,
    Pennsylvania State University,
    University Park, PA 16802, USA
}
\author{Fei~Teng}
\affiliation{
    Institute for Gravitation and the Cosmos,
    Pennsylvania State University,
    University Park, PA 16802, USA
 }

\date{\today}

\begin{abstract}

We revisit the quantum-amplitude-based derivation of the gravitational waveform emitted by the scattering of two spinless massive bodies at the third order in Newton's constant, $h \sim G+G^2+G^3$ (one-loop level), and correspondingly update its comparison with its classically-derived multipolar-post-Minkowskian counterpart. A spurious-pole-free reorganization of the one-loop five-point amplitude substantially simplifies the post-Newtonian expansion. We find complete agreement between the two results up to the fifth order in the small velocity expansion after taking into account three subtle aspects of the amplitude derivation: (1) in agreement with [arXiv:2312.07452 [hep-th]], the term quadratic in the amplitude in the observable-based formalism [JHEP \textbf{02}, 137 (2019)] generates a frame rotation by half the classical scattering angle; (2) the dimensional regularization of the infrared divergences of the amplitude introduces an additional $(d-4)/(d-4)$ finite term; and (3) zero-frequency gravitons are found to contribute additional terms both at order $h \sim G^1$ and at order $h \sim G^3$ when including disconnected diagrams in the observable-based formalism.

\end{abstract}

\maketitle



\section{Introduction}

The remarkable detection of gravitational waves by the LIGO and Virgo collaborations~\cite{LIGOScientific:2016aoc,LIGOScientific:2017vwq} sparked the development of novel quantum field theory (QFT)-based approaches to the general relativistic two-body problem which build on enormous advances in scattering amplitudes and effective field theory. The scattering regime at large minimum separation is a theoretical environment free of many subtleties related to the definition of initial and final states, making it the ideal ground for thorough comparison and cooperation with the well-tested traditional analytical approaches to this problem. 
In this paper we revisit such a comparison between the amplitudes-based prediction in the observable-based (KMOC) formalism~\cite{Kosower:2018adc, Cristofoli:2021vyo} for the classical gravitational wave signal from the scattering of two non-spinning masses and the corresponding waveform derived within the Multipolar-post-Minkowskian (MPM) formalism~\cite{Blanchet:1985sp,Blanchet:1989ki,Blanchet:2013haa}.

A natural framework for discussing classical scattering from the QFT perspective consists in taking the classical limit ($\hbar \to 0$) of the perturbative QFT expansion of scattering amplitudes, which is an expansion in powers of $\frac{G m_1 m_2}{J c}$, where $J$ is the orbital angular momentum. By collecting the appropriate terms contributing to classical observables, one should obtain  the classical post-Minkowskian (PM) expansion (in powers of  Newton's constant $G$) of the considered classical observable (e.g. an impulse, $\Delta p^\mu $, or the waveform $h_{\mu\nu} \equiv  g_{\mu\nu} - \eta_{\mu\nu}$). For pioneering computations of the classical PM expansion of scattering observables, see \cite{Peters:1970mx,Kovacs:1977uw} (waveform) and \cite{Portilla:1980uz,Westpfahl:1985tsl} (impulse). Classical bodies are treated as point particles with additional properties encoded in higher-dimension operators extending their minimal interactions with gravity. The classical limit is identified in the spirit of the correspondence principle, and Lorentz invariance of standard QFT perturbation theory facilitates keeping the complete dependence on the velocity of particles at every order in the $G$ expansion. 

Extensive effort, utilizing advances in scattering amplitude methods such as of generalized unitarity~\cite{UnitarityMethod,BernMorgan,Fusing,Bern:1997sc,Britto:2004nc},  the double-copy relations~\cite{BCJ,BCJLoop,KLT} and powerful integration methods~\cite{Chetyrkin:1981qh,Laporta:2000dsw,Smirnov:2008iw,Smirnov:2019qkx,Kotikov:1990kg,Bern:1993kr,Remiddi:1997ny,Gehrmann:1999as} and complemented by worldline approaches~\cite{Mogull:2020sak,Kalin:2020mvi}, 
explored and evaluated inclusive  scattering observables for interactions of spinning and spinless bodies, such as scattering angles, impulse, etc., to remarkably high perturbative orders~\cite{Bern:2021dqo, Bern:2021yeh,Dlapa:2021npj,Dlapa:2021vgp,Dlapa:2022lmu,Jakobsen:2023ndj,Jakobsen:2023hig}. 
Similarly, position-dependent (point-like) observables such as the gravitational waveform emitted during a weak-field scattering process have been discussed from different perspectives and approaches~\cite{Cristofoli:2021vyo, Jakobsen:2021smu,Mougiakakos:2022sic, Riva:2022fru}. 
Recently, aspects of radiation and reaction were discussed in KMOC formalism at the third order in Newton's constant in \cite{Elkhidir:2023dco} and the scattering waveform, $h_{\mu\nu} \sim G^1+G^2+G^3+\dots$, was constructed to this order for spinless particles~\cite{Herderschee:2023fxh,Georgoudis:2023lgf,Brandhuber:2023hhy} (see also \cite{Caron-Huot:2023vxl}), for spinning particles at $O(G^2)$~\cite{DeAngelis:2023lvf,Brandhuber:2023hhl,Aoude:2023dui} and at $O(G^3)$ \cite{Bohnenblust:2023qmy} and various orders in spin.

Gravitational wave emission from gravitating binaries has been studied with traditional analytic methods over many years, with the purpose of constructing analytical-based waveforms to be used in the detection and data-analysis of gravitational wave signals. One of the most successful analytical methods for computing the GW emission from generic sources has been the (post-Newtonian-matched) Multipolar-Post-Minkowskian (MPM) formalism~\cite{Blanchet:1985sp,Blanchet:1989ki,Blanchet:2013haa}. The MPM formalism has been developed over the years to a high perturbative accuracy, recently reaching the fourth post-Newtonian (4PN) accuracy~\cite{Blanchet:2023bwj,Blanchet:2023sbv}, i.e.  the N$^4$LO level in a PN expansion (in powers of $\frac{v^2}{c^2}+ \frac{G m}{r c^2}$) beyond the leading-order quadrupole formula pioneered by Einstein~\cite{Einstein:1918btx}. This PN accuracy includes nonlinear terms of 
 order $O(G^5)$ in $h_{\mu\nu}$. 

A comparison of the MPM and amplitudes-based results \cite{Bini:2023fiz} revealed substantial differences. Ref.~\cite{Bini:2023fiz} also pointed out a remarkable structure: if the amplitudes-based results are interpreted as being in a frame rotated by half the classical scattering angle, then the differences are dramatically reduced.
The origin of this frame rotation was recently understood by  Georgoudis, Heissenberg and Russo~\cite{Georgoudis:2023eke} in terms of a certain cut contribution pointed out in Ref.~\cite{Caron-Huot:2023vxl}, and demonstrated explicitly in a small-frequency expansion.\footnote{In the version~2 update of~\cite{Georgoudis:2023eke}, this relation between the cut contribution and frame rotation was also tested to NNLO (1PN) in the PN expansion.} 

In this paper we revisit several aspects of the amplitude- and observable-based (KMOC) \cite{Cristofoli:2021vyo} derivation of the classical gravitational wave signal emitted during the scattering of two non-spinning masses at the NNLO order,\footnote{In the following we will mainly refer to this as ``the EFT waveform'' following the nomenclature used in the original comparison of \cite{Bini:2023fiz}, but at times we will also use ``the amplitudes-based waveform'' and the ``KMOC waveform''.} $h_{\mu\nu} =  G^1+G^2+G^3$, and of the comparison of the frequency-domain result with the corresponding waveform derived within the  Multipolar-post-Minkowskian formalism. We find complete agreement to the available post-Newtonian accuracy (see Secs.~\ref{sec:Comparison}, \ref{sec:epBYepANDnu} and \ref{sec:discANDnusq}).

The main difficulty with carrying out the PN expansion of the amplitude-based scattering waveform stems from the severe spurious singularities exhibited by the available results~\cite{Herderschee:2023fxh,Brandhuber:2023hhy,Georgoudis:2023lgf}. 
We reorganize both the amplitude and the cut expressions so that they are written in terms of functions free of such spurious poles on the physical sheet (see \cref{app:reorg}). This facilitates the PN expansion, which we carry out here to 3PN accuracy\footnote{Ref.~\cite{Bini:2023fiz} had reached the 2.5PN accuracy for the $\phi$-even part of the waveform, without incorporating the 2PN-level terms.}, in the equatorial ($\theta = \pi/2$) plane and for the complete dependence on the azimuthal angle $\phi$. We demonstrate that the frame rotation of \cite{Bini:2023fiz} is equivalent to the connected part of the cut contribution of the KMOC formalism through this order~(see \cref{sec:cut_and_rotation}). 

Gravitons of vanishing frequency and their contributions to observables are a notoriously thorny subject, see e.g.~\cite{Veneziano:2022zwh}. Ignoring them in the KMOC calculation leads to differences with the MPM results already at leading order in Newton's constant (in the time-independent Coulombic field $h_{\mu\nu}=O(G) $ generated by the incoming particles). We will show that including them resolves this difference (see Sec.~\ref{sec:constant_background}). They also have nontrivial contributions at higher orders in Newton's constant, where, through the disconnected cut contributions, they yield terms that can be interpreted as a Bondi-Metzner-Sachs (BMS)~\cite{Bondi:1962px, Sachs:1962wk} supertranslation of the result with no contributions from such states (see Secs. \ref{sec:discANDnusq} and \ref{sec:BMSgeneral}). This BMS supertranslation is the same one that clarifies~\cite{Veneziano:2022zwh} the existence of an angular momentum loss at $O(G^2)$~\cite{Damour:2020tta}; its importance in the amplitudes-MPM comparison was initially observed in a low-frequency expansion at the first non-universal order~\cite{Georgoudis:2023eke}.
A last ingredient in our comparison is a subtle $O(\epsilon/\epsilon)$ contribution which originates from the graviton being described by a symmetric, traceless $(d-2) \times (d-2)$ matrix in $d$ dimensions (see Secs.~\ref{ampANDconnectedcuts} and \ref{sec:epBYepANDnu}) in our (retarded-time renormalization) scheme.

We begin with a brief description of the MPM and amplitudes-based methods used to evaluate the waveforms, emphasizing the new contributions in the latter, and then proceed to detaling our comparison. We collect the list of notations used in this paper in \cref{app:notation}.

\paragraph*{Note added:} While this paper was being written up, we became aware of concurrent work by Georgoudis, Heissenberg and Russo~\cite{Georgoudis:2024pdz}, which partly overlaps with aspects of our analysis. We thank the authors for communication and coordination on the submission.

\section{Introducing the Multipolar Post-Minkowskian (MPM) formalism}
\label{sec:IntroMPM}

The two transverse-traceless components 
\be
\label{eq:asym_metric}
f_+=\lim_{r\to \infty}(r\,  h_+)\,,\qquad f_\times=\lim_{r\to \infty}(r\,  h_\times)
\ee
of the classical, time-domain asymptotic waveform are  encoded in the complex quantity
\begin{align}
\label{eq:asym_waveform}
W(T_r,\theta,\phi) &= \frac{c^4}{4 G} (f_+- i f_\times) \nn \\
&= \frac{c^4}{4 G} \lim_{r\to \infty} \mb^{\mu } \mb^{\nu }r\, h_{\mu \nu}\,,
\end{align}
where the normalization prefactor $\frac{c^4}{4 G}$ is used to simplify the PN expansion of $W$.\footnote{We indicate powers of $c$, or $\eta\equiv \frac1c $, when they help to understand the PN order of various quantities. Otherwise, we often set $c$ to 1 without warning.} 
Here  $T_r \simeq t- \frac{r}{c} - 2 \frac{G \ME}{c^3} \log \frac{r}{c b_0}$ denotes a retarded time which contains a logarithmic shift, involving an  arbitrary time scale $ b_0$, proportional to the total center-of-mass (c.m.) mass-energy of the system,  $ \ME \equiv \frac{E}{c^2}$, while  $\mb^{\mu }$ (defined below) is a complex, null polarization vector tangent to the sphere at infinity.

The MPM approach is set up in the c.m. of the system, i.e. with time axis defined in terms of the incoming momenta,
$p_a, (a = 1,2)$, as
\be
\label{eq:e0}
{e}_0^\mu= \frac{{ p}_1^\mu+{ p}_2^\mu}{|{  p}_1^\mu+{ p}_2^\mu|}\,.
\ee
At the present PM accuracy of the EFT waveform  we can equivalently define the c.m. system by choosing as unit time-vector
\be
\label{eq:ebar0}
{\bar e}_0^\mu= \frac{{\bar p}_1^\mu+{\bar p}_2^\mu}{|{\bar  p}_1^\mu+{\bar p}_2^\mu|}\,,
\ee
where ${\bar p}_a$ denote  the {\it classical} averaged momenta, namely
\be 
\label{eq:pbar}
\pb_a=\frac12 (p_a+p'_a)\,,\quad (a = 1,2)\,.
\ee 
Here, $p'_a$ are the classical outgoing momenta, differing from the incoming ones $p_a$ by the full classical scattering angle $\chi = O(G)$. However, ${\bar e}_0^\mu$ differs from ${e}_0^\mu$ only at $O(G^3)$,  which is the order of
the radiated momentum $P^{\rm rad}=-(p'_1 + p'_2-p_1 -p_2)$.

It is convenient, in the MPM formalism, to compute
 the (spatial) components of the radiative multipole moments of the binary system with respect to a spacetime frame 
 ${\bar e}_0$, $e_x$, $e_y$, $e_z$ anchored on the averaged momenta  $\pb_a$ rather than on the
 incoming ones $p_a$.
The vector $e_y$ lies in the spatial direction of ${\bar p}_1$ 
%
%
%
%
(i.e.  the bisector between the incoming and the outgoing spatial momentum of the first particle in the \com frame),
\be
\label{eq:p1p2}
{\bar p}_1 = {E}_1  {\bar e}_0 +  {\bar P}_{\rm c.m.} e_y\,,\qquad {\bar p}_2 = {E}_2  {\bar e}_0 -  {\bar P}_{\rm c.m.} e_y\,.
\ee
Here, $E_a = \sqrt{m_a^2+P_{\text{c.m.}}^2}$, and $P_{\text{c.m.}}$ is the magnitude of the spatial part of the incoming momenta. In principle, ${\bar P}_{\text{c.m.}}$ differs from $P_{\text{c.m.}}$ by ${\bar P}_{\rm c.m.} = {P}_{\rm c.m.}  \cos \frac12 \chi$. However, the difference is only at the order $\chi^2 = O(G^2)$ and will be negligible in our work.
The axis vector $e_x$ is in the plane of motion and orthogonal to $e_y$ (and oriented from particle 2 towards
particle 1). It is also the direction of the relative impact parameter,
\be
\label{eq:bbar}
b_{12} = b\, e_x\,.
\ee
Here $b$ is an eikonal-type impact parameter, linked to the incoming-momenta impact parameter $b_{\text{in}}$ by
$b_{\text{in}} = b \,\cos\frac{1}{2}\chi$, in which the difference $O(\chi^2)=O(G^2)$ is again negligible.
The last spatial axis vector $e_z$ is orthogonal to the plane of motion (and such that $e_x$, $e_y$, $e_z$
is positively oriented).
All vectors and tensors are decomposed in the frame  ${\bar e}_0$, $e_x$, $e_y$, $e_z$ and
the angles $\theta,\phi$ are accordingly defined, so that
\begin{align}
{\bf n}(\theta,\phi) &= \sin\theta \cos\phi \, e_x +\sin\theta\sin\phi \, e_y +\cos\theta \,e_z \,, \nn\\
k &= \om \left( {\bar e}_0 + {\bf n}(\theta,\phi) \right), \nn \\
\mb &= \frac{1}{\sqrt{2}} \left[\D_\theta {\bf n}(\theta,\phi)-  \frac{i}{\sin \theta} \D_\phi {\bf n}(\theta,\phi)\right].
\label{frame_exey}
\end{align}
Here ${\bf n}$ is the spatial unit vector that characterizes the direction of the gravitational wave propagation. 
%

 The Multipolar post-Minkowskian (MPM) formalism \cite{Blanchet:1985sp,Blanchet:1989ki,Blanchet:2013haa}
 computes the time-domain waveform $W^{\rm MPM}(T_r,\theta,\phi)$ as a sum over
 irreducible multipolar contributions, keyed by their multipole order $\ell$ and their spatial parity. Even-parity
 (radiative) multipoles are denoted $U_\ell$, $\ell=2,3,4,\ldots$, while odd-parity ones are denoted 
 $V_\ell$, $\ell=2,3,4,\ldots$. It is convenient to keep track of the $\eta \equiv \frac1c$ prefactors 
 entering the multipole expansion of the waveform emitted by a slow-motion source by writing $W$ as 
 \bea \label{WMPM1}
W^{\rm MPM}(T_r,\theta,\phi)&=& U_2+ \eta (V_2 +U_3) + \eta^2 (V_3+U_4)\nonumber\\ 
&+& \eta^3 (V_4+U_5)+  \eta^4 (V_5+U_6) \nn\\ &+& \eta^5 (V_6+U_7)+ \cdots \,.
\eea
With this normalization, the PN expansion of each radiative multipole $U_\ell$ or $V_\ell$ starts at
the Newtonian order (in the sense that the leading order (LO) value of each multipole is expressed in terms of
Newtonian-level quantities). In Eq. \eq{WMPM1}, each $U_\ell$ or $V_\ell$ is expressed in terms
of corresponding symmetric-trace-free (STF) Cartesian tensors of order $\ell$, according to
\bea
U_\ell(T_r,\theta,\phi) &=& \frac{1}{\ell!} \mb^{i} \mb^{j } n^{i_1} n^{i_2} \cdots n^{i_{\ell-2}} U_{i j i_1 i_2 \cdots i_{\ell-2}}(T_r)\,, \nn\\
V_\ell(T_r,\theta,\phi) &=& - \frac{1}{\ell!}\frac{2\ell}{\ell+1}\mb^{i} \mb^{j } n^c  n^{i_1} n^{i_2} \cdots n^{i_{\ell-2}} \nonumber\\
&&  \times\;\epsilon_{cd i}V_{j d i_1 i_2 \cdots i_{\ell-2}}(T_r)\,.
\eea
The two infinite sequences of {\it radiative multipole moments} $U_{i_1 i_2 \cdots i_{\ell}}(t)$,  $V_{i_1 i_2 \cdots i_{\ell}}(t)$
(where $t$ henceforth denotes the retarded time variable $T_r$) fully, and uniquely, parametrize the time, 
and angular, dependences of any gravitational waveform. They are formally observable if we imagine
surrounding the emitting system by infinitely many GW detectors located on the sphere at infinity  (in the c.m. frame). 
Actually, only the time-dependent part of each multipole, e.g., $ {U}_{i_1 i_2 \cdots i_{\ell}}(t) - { U}_{i_1 i_2 \cdots i_{\ell}}(t=-\infty)  $, is directly observable when considering the LO $O(\frac1r)$ asymptotic gravitational field at future null infinity.
Perturbative computations have, however, shown that the presence of non-zero initial values ${ U}_{i_1 i_2 \cdots i_{\ell}}(t=-\infty) \neq 0$ of the radiative multipoles have physical effects
at future null infinity at higher orders in $O(\frac1r)$, via the back-scattering of gravitational waves on spacetime curvature (``tails'') \cite{
Damour:1985cm,Kehrberger:2024clh}.
We will come back to this issue below. 

The (post-Newtonian(PN)-matched) MPM formalism computes each (time-domain)  radiative multipole moment
$U_{i_1 i_2 \cdots i_{\ell}}(t)$,  $V_{i_1 i_2 \cdots i_{\ell}}(t)$ in terms of the stress-energy tensor of the 
material source, $T^{\mu \nu}(x^\lambda)$,  in several steps. At the end of the day, each radiative multipole at
retarded time $t$ is given by a sum of contributions involving both source variables at time $t$
and hereditary integrals over the past behavior of the source, i.e. over times $ -\infty \leq t' \leq t$.
All those hereditary integrals are expressed in terms of some {\it source multipole moments}, of mass-type,
$I_{i_1 i_2 \cdots i_{\ell}}(t)$,  and spin-type, $J_{i_1 i_2 \cdots i_{\ell}}(t)$. Some of the hereditary integrals are linear
in the source moments (``linear tails'') while others are nonlinear because they arise from nonlinear couplings
between the multipole signals in the zone exterior to the source. Finally, the source moments are computed in terms
of the dynamical variables of the source by using a PN-expanded approach to the gravitational field in the near zone.

To give an idea of the structure of the expressions coming out of the MPM formalism, let us display a few of the
contributions to the radiative quadrupole\footnote{In the formulas below the angular brackets $\langle\dots\rangle$ denote
a STF projection over the enclosed indices.} $U_{ij}(t)$ at the fractional 2.5PN accuracy ($\eta^5$ beyond the 
classic Einstein quadrupole formula):
\bea \label{UijMPM0}
U_{ij}(t) &=& \frac{d^2  I_{ij}(t)}{dt^2} \\
&+& \frac{2G \ME}{c^3}\int_0^\infty d\tau I_{ij}^{(4)}(t-\tau) \left(\log\left(\frac{\tau}{2 b_0}\right)+ \frac{11}{12} \right) \nn \\
&+& \frac{G}{c^5}\left(\frac17 I_{a\langle i}^{(5)}I_{j\rangle a}^{\vphantom{(5)}}-\frac57 I_{a\langle i}^{(4)}I_{j\rangle a}^{(1)}-\frac27 I_{a\langle i}^{(3)}I_{j\rangle a}^{(2)}\right) + \cdots \nonumber
\eea
in which $\ME \equiv \frac{E}{c^2}$ is the total mass-energy of the system, and one must insert the 2.5PN-accurate value of the source quadrupole $ I_{ij}(t)$. 

In the MPM formalism, the successive contributions indicated in Eq.~\eqref{UijMPM0} have different origins:
(i) the source quadrupole moment $I_{ij}(t)$ computes (in a PN-expanded approximation) the quadrupole moment of the effective stress-energy tensor of the system, i.e. the sum of the $T^{\mu\nu}_{\rm pp}$ of the point particles and of the effective $T^{\mu\nu}_{\rm g}$ of the near-zone gravitational field (potential gravitons, plus some local retarded-field effects); (ii) the second (hereditary) term is the linear tail contribution generated by the back-scattering of quadrupolar waves on the external-zone Coulomb field $\sim \frac{ G \ME}{r}$  generated by the total mass-energy of the system; and (iii) is an example of the several (instantaneous) nonlinear terms arising from  cubic couplings between multipole moments (here quadrupole $\times$ quadrupole $\times$ quadrupole coupling). 

In the \com frame the 2.5PN-accurate  source quadrupole $I_{ij}$
 has the following structure 
\bea \label{IijPN0}
I_{ij}^{\leq 2.5PN}(t) &=& \nu M \left( 1 + \frac{a_2}{c^2} +  \frac{a_4}{c^4} - \frac{24}{7} \frac{\nu}{c^5} \frac{G^2 M^2}{r^2} \dot r\right) x^{\langle i} x^{j \rangle} \nn \\
&+& {\rm terms \; in } \;  x^{\langle i} v^{j \rangle} \; {\rm and} \;  v^{\langle i} v^{j \rangle}\, .
\eea 
Here, $x^i(t)= x_1^i(t)- x_2^i(t)$ denotes the relative motion (in the \com frame), $v^i(t) = \frac{d x^i}{dt}$ is the relative velocity, $M=m_1+m_2$ is the total mass and $\nu$ denotes the symmetric mass ratio, $\nu=m_1 m_2/(m_1+m_2)^2$.
In addition, in order to get explicit results for the radiative moments as a function of time, one must solve the
correspondingly PN-accurate equations of motion of the binary system. The latter task is aided by the existence
of a rather simple quasi-Keplerian representation of the hyperbolic-like solution of the 2PN-level conservative
equations of motion (to which the effect of radiation reaction, which starts at the $G^2/c^5$ level, must be added) \cite{DD85,Damour:1988mr,Cho:2018upo}.

As the asymptotic behavior of  $x^i(t)$ in the incoming state is $x^i(t)  \approx v^i_0 \, t + O(\log t)$, the source
quadrupole moment is found to behave as $I_{ij}(t) \approx  \nu M  \text{STF}[ v^i_0  v^j_0] \left(1 + O(\frac{v_0^2}{c^2})\right) t^2$.
One then checks that all the hereditary integrals defining $U_{ij}(t)$ are convergent, and that  $U_{ij}(t)$ has
a {\it nonzero} value when $t \to -\infty$ of the form
\be
U_{ij}(t= - \infty) = 2 \nu M \, \text{STF}[ v^i_0  v^j_0] \left(1 + O(\frac{v_0^2}{c^2}) \right)\,.
\ee
Note that the nonzero value of the radiative quadrupole moment at the Newtonian approximation
is a simple consequence of the classic Einstein quadrupole formula saying that 
$U_{ij}  = \frac{d^2}{dt^2} (\sum_a  \text{STF}(m_a x_a^i x_a^j)) + O(\eta^2)$.

Similar results hold for all the multipole moments. Indeed, the origin of all these nonzero initial radiative
multipole moments is the fact that the waveform in the infinite past is simply given by the
Coulombic field of the incoming worldlines, namely
\be \label{fin}
f_{ij}(t= - \infty,\n)= \sum_a 4 G   \left[  \frac{\left(p_a^i p_a^j\right)^{\rm TT}}{E_a- \n \cdot \p_a} \right],
\ee
leading to (with $ n\equiv k/\om$)
\be \label{Win}
W(t= - \infty) =  \sum_a \frac{ (\mb \cdot p_a)^2}{E_a- \n \cdot \p_a}=  \sum_a \frac{ (\mb \cdot p_a)^2}{- ( n\cdot p_a)}\,.
\ee
In other words, the MPM waveform necessarily incorporates the nonzero initial value, Eqs. \eq{fin}, \eq{Win}, given by the Coulombic  field of the incoming worldlines.

Finally, after having obtained some representation of the various time-domain multipole moments, one must compute
their Fourier transforms (over the retarded time variable), namely
\bea
U_{i_1 i_2 \cdots i_{\ell}}(\om)= \int_{- \infty}^{+ \infty} dt e^{i \om t} U_{i_1 i_2 \cdots i_{\ell}}(t)\,, \nn \\
V_{i_1 i_2 \cdots i_{\ell}}(\om)= \int_{- \infty}^{+ \infty} dt e^{i \om t} V_{i_1 i_2 \cdots i_{\ell}}(t)\,.
\eea
This leads to a Fourier-domain MPM waveform of the form
\bea
\label{Wom_th_phi}
W^{\rm MPM}(\om,  \theta,\phi) &\equiv& U_2(\om,\theta,\phi)\\
&+& \eta (V_2(\om,\theta,\phi) +U_3(\om,\theta,\phi))\nonumber\\ 
&+& \eta^2 (V_3(\om,\theta,\phi)+U_4(\om,\theta,\phi)) \nonumber\\ 
&+& \eta^3 (V_4(\om,\theta,\phi)+U_5(\om,\theta,\phi)) \nonumber\\ 
&+&  \eta^4 (V_5(\om,\theta,\phi)+U_6(\om,\theta,\phi)) \nn\\ &+& \eta^5 (V_6(\om,\theta,\phi)+U_7(\om,\theta,\phi))
+ \cdots  \nonumber
\eea
Here, the explicit powers of $\eta = \frac1c$ indicate those associated with the LO, Newtonian
value of each radiative multipole. They serve as a reminder that the knowledge of the full waveform
$W^{\rm MPM}(\om,  \theta,\phi)$ at, say, the 2.5PN accuracy  requires one to know: 
(i) $U_2$ at the fractional 2.5PN accuracy ($\eta^5$); 
(ii) $V_2$ and $U_3$ at the fractional 2PN accuracy ($\eta^4$);
(iii) $V_3$ and $U_4$ at the fractional 1.5PN accuracy ($\eta^3$); 
(iv) $V_4$ and $U_5$ at the fractional 1PN accuracy ($\eta^2$);
(v) $V_5$ and $U_6$ at the Newtonian accuracy (because their first PN correction starts at 1PN) and, finally,
(vi) $V_6$ and $U_7$ at the Newtonian accuracy.

In addition, as we are comparing the MPM waveform to the one-loop-accuracy amplitudes-based waveform, we only need
to determine the $O(G)$ and $O(G^2)$ parts of the $G-$expansion of the waveform (in its $W$ guise, ie.
after dividing $\lim_{r\to \infty} \mb^{\mu } \mb^{\nu }r\, h_{\mu \nu}$ by $4 G$). 
Finally, an MPM/EFT waveform comparison at the 2.5PN accuracy requires, on the MPM side, the
knowledge of:
\bea 
U_2(\om) &\sim& (G+G^2)(\eta^0+ \eta^2 +\eta^3+\eta^4+\eta^5) \nn \\
V_2(\om) \; {\rm and} \; U_3(\om) &\sim& (G+G^2)(\eta^0+ \eta^2 +\eta^3+\eta^4) \nn \\
V_3(\om) \; {\rm and}\; U_4(\om) &\sim& (G+G^2)(\eta^0+ \eta^2 +\eta^3) \nn \\
V_4(\om) \; {\rm and}\; U_5(\om) &\sim& (G+G^2)(\eta^0+ \eta^2 ) \nn \\
V_5(\om) \; {\rm and}\; U_6(\om) &\sim& (G+G^2)(\eta^0) \nn \\
V_6(\om) \; {\rm and}\; U_7(\om) &\sim& (G+G^2)(\eta^0)\,.
\eea
The time-dependent part ($\om \neq 0$)
 of each multipole starts at order $G^1$. Only their static (initial) values contribute at order $G^0$, as exhibited 
 in Eq. \eqref{Win}.
 
In these expansions, the terms of order $\eta^3$ only arise at order $G^2$ and have two different origins.
One origin is  the linear tails which are all of the form
\begin{align}
U_\ell^{G^2\rm tail}(t) &=  \frac{2G \ME }{c^3}\int_0^\infty d\tau \frac{d^2}{dt^2 }U_\ell^{G}(t-\tau) \nonumber\\
&\hspace{2cm}\times\left(\log\left(\frac{\tau}{2 b_0}\right)+ \kappa_\ell \right) \nn \\
V_\ell^{G^2\rm tail}(t) &=  \frac{2G \ME }{c^3}\int_0^\infty d\tau \frac{d^2}{dt^2 }V_\ell^{G}(t-\tau) \nonumber\\
&\hspace{2cm}\times\left(\log\left(\frac{\tau}{2 b_0}\right)+ \sigma_\ell \right), 
\end{align}
with
\bea
\kappa_\ell&=&\frac{ 2 \ell^2+5 \ell+4}{\ell(\ell+1)(\ell+2)} + \sum_{k=1}^{k=\ell-2} \frac1k ,\nn \\
 \sigma_\ell&=&\frac{\ell-1}{\ell(\ell+1)} + \sum_{k=1}^{k=\ell-1} \frac1k ,
\eea
in the time domain, and
\bea
U_\ell^{G^2\rm tail}(\om) &=&  \frac{2G \ME \om}{c^3}\left(\frac{\pi}{2}+ i \left[\log(2 \om b_0 e^{\gamma_E})- \kappa_\ell\right] \right)U_\ell^{G}(\om), \nn \\
V_\ell^{G^2\rm tail}(\om) &=&   \frac{2G \ME \om}{c^3}\left(\frac{\pi}{2}+ i \left[\log(2 \om b_0 e^{\gamma_E})- \sigma_\ell \right]\right)V_\ell^{G}(\om), \nn\\
\eea
in the frequency domain.

Another origin of the $G^2 \eta^3$ terms exists for $V_3$ and $U_4$. They come from quadrupole$\times$quadrupole
couplings. E.g., in $U_4$ there are terms of the form
\bea
U_{ijkl}^{\rm II}(t)&=&\frac{G}{c^3}\left(-\frac{21}{5}I^{\vphantom{(5)}}_{\langle ij}I_{kl\rangle}^{(5)}-\frac{63}{5}I_{\langle ij}^{(1)}I_{kl\rangle}^{(4)}\right.\nonumber\\
&-&\left. \frac{102}{5}I_{\langle ij}^{(2)}I_{kl\rangle}^{(3)}
\right)\,.
\eea
These quadrupole$\times$quadrupole couplings generate terms of order $G^2/c^5$ in $U_2$ (see last line
in Eq. \eq{UijMPM0}).

On the other hand, the terms of even order in $\eta$ come from PN corrections to the source multipole moments,
as exemplified by the terms $ \frac{a_2}{c^2} +  \frac{a_4}{c^4}$ in Eq.~\eq{IijPN0}.

\section{Structure and computation of the MPM waveform}
\label{sec:MPMwaveform}

Ref. \cite{Bini:2023fiz}  computed the multipoles needed to reach the absolute $(G^1+G^2) \eta^5$ accuracy in the even-in-$\phi$ projection of the waveform.  Because of the planar nature (in the $ x y$ plane) of the binary motion generating the waveform, it is easily seen that a parity-even multipole $U_l$ contains (when decomposed in $e^{i m \phi}$) $m$ values 
equal to $m=l, l-2, l-4$ etc. By contrast, a parity-odd multipole $V_l$ contains $e^{i m \phi}$ contributions for
$m=l-1, l-3, l-5$ etc.  Therefore, the even-in-$\phi$ projection of $W^{\rm MPM}$ is given, at accuracy $\eta^5$, by $U_2$, $V_3$, $U_4$, $V_5$, and $U_6$. The latter multipoles were computed in Ref. \cite{Bini:2023fiz}, and their
values in the equatorial plane ($\theta=\frac{\pi}{2}$) were displayed there. 

In the present work,
we complete the MPM/EFT comparison by also computing some of the multipoles contributing to the  odd-in-$\phi$ part
of the waveform. For simplicity, we focus on only two terms in the PN expansion of the odd-in-$\phi$ piece,
namely, the terms of order $(G^1+G^2) (\eta^1+ \eta^4)$. These terms come from the Newtonian-level terms
in $V_2$ and  $U_3$, together with the corresponding tail terms (which are $G \eta^3$ smaller than the LO contribution).
We present the full angular dependence of the multipoles by giving the coefficients of the
decomposition of $W^{\rm MPM}(\om)$ in spin-weighted (with spin-weight $s=-2$) spherical harmonics (SWSH)
${}_{s}Y_{lm}(\theta,\phi)$, defined with the convention spelled out in an ancillary file \texttt{MPMmultipoles.nb}. For instance, denoting
for clarity $s=-2$ as $\bar 2$, and $m=-2$ as $\bar 2$)
\bea
{}_{{\bar 2}}Y_{22}(\theta,\phi) &=& \sqrt{\frac{5}{ 4\pi}} e^{2 i \phi}  \cos^4\left(\frac{\theta}{2}\right), \nn \\
{}_{{\bar 2}}Y_{2 {\bar 2}}(\theta,\phi) &=& \sqrt{\frac{5}{ 4\pi}}  e^{-2 i \phi} \cos^4\left(\frac{\theta}{2}\right)\,.
\eea
The $(l,m)$ SWSH coefficients of $W^{\rm MPM}$ are the sum of the SWSH coefficients of  $U_l$ and  $V_l$ (respectively
multiplied by $\eta^{l-2}$ and   $\eta^{l-1}$), 
\be
W^{\rm MPM}_{lm}= \eta^{l-2} U_{lm}+ \eta^{l-1} V_{lm}\,,
\ee
where the coefficients $ U_{lm},  V_{lm}$ are defined by
\bea
U_l(\theta,\phi) &=&\sum_{m=-l}^l U_{lm}\, {}_{\bar 2}Y_{lm}(\theta,\phi), \nn \\
V_l(\theta,\phi) &=&\sum_{m=-l}^l V_{lm}\, {}_{\bar 2}Y_{lm}(\theta,\phi).
\eea
The coefficients $ U_{lm},  V_{lm}$ can be obtained by using the orthonormality of the ${}_{s}Y_{lm}(\theta,\phi)$'s, e.g.
\beq
U_{lm}=\int \sin\theta d\theta d\phi \, U_l(\theta,\phi)\;  {}_{\bar 2}Y_{2,m}^*(\theta,\phi)\,.
\eeq
In the time-domain, the  $ U_{lm}$, and $ V_{lm}$ are functions of the retarded time $t$. When going to
the frequency domain, they are functions of $\om$, with, e.g., 
\be
U_{lm}(\om) = \int_{- \infty}^{+ \infty} dt e^{i \om t} \, U_{lm}(t)\,.
\ee
In practice, the Fourier transforms of the radiative multipoles are more conveniently performed
on the Cartesian representation of  $U_{ i_1 i_2 \cdots i_{l}}(t)$ and  $V_{ i_1 i_2 \cdots i_{l}}(t)$.
As explained in Ref. \cite{Bini:2023fiz}, it is convenient to Fourier transform the first time-derivatives
of each  $U_{ i_1 i_2 \cdots i_{l}}(t)$ and  $V_{ i_1 i_2 \cdots i_{l}}(t)$ (which vanish like $O(1/t^2)$
at $t \to -\infty$) and to divide the result by $- i \om$. We only consider strictly positive frequencies $\om >0$
because the $\om <0$ part is determined by the reality of  $U_{ i_1 i_2 \cdots i_{l}}(t)$ and  $V_{ i_1 i_2 \cdots i_{l}}(t)$,
while the $\om=0$ part is fully described by the ingoing, asymptotic waveform $W(t=- \infty)$. $W(t=- \infty)$ is exactly given by the $G^0$, linearized-gravity result Eq. \eqref{Win}.

Let us illustrate the structure (and physical content) of   $ U_{lm}(\om)$, and $ V_{lm}(\om)$, by focussing 
on the quadrupolar contribution $U_2 = \frac12 \mb^i \mb^j U_{ij}$ which is the only one which starts at the
Newtonian order $\eta^0$. 

We have sketched in Eq.~\eqref{UijMPM0} the various contributions to $U_{ij}(t)$ which contribute at the
fractional $\eta^5$ (2.5PN) level. The first term in Eq.~\eqref{UijMPM0}, 
\be
U_{ij}^{I}(t) =\frac{d^2  I_{ij}(t)}{dt^2}\,,
\ee
has $\eta^5$ contributions coming both from an explicit 2.5PN contribution in the source
quadrupole moment \cite{Blanchet:1996wx},
\be \label{U2I}
 I_{ij}(t) \sim \text{STF}\Big[\sum_a m_a x_a^i x_a^j + \eta^2+ \eta^4+ \eta^5 +\ldots\Big]\,,
 \ee
 from the need to include $O(G^2/c^5)$ radiation-reaction effects \cite{Damour:1981bh} in the second time-derivative of the positions $x_a^i(t)$, 
 and from the need to compute the Fourier transform along the radiation-reacted hyperbolic motions\footnote{
 However, as explained in Ref.~\cite{Bini:2023fiz}, this can be avoided by working with $\frac{d}{dt} U_{ij}^{I}(t) $.}.
 
 The second term in Eq. \eqref{UijMPM0} is the (linear) tail contribution which starts at order $G^2 \eta^3$
 and reads, in the frequency domain, 
\begin{align}\label{U2tail}
    U_{ij}^{G^2\rm tail}(\om) &= \frac{2G {\sf M} \om}{c^3} \\
    &\quad \times\left[\frac{\pi}{2}+ i \left(\log(2 \om b_0 e^{\gamma_E})-\frac{11}{12} \right) \right] U_{ij}^{G}(\om)\,,\nonumber
\end{align}
where we recall that ${\sf M} \equiv \frac{E}{c^2}$. Here $b_0$ is the time scale used in the MPM formalism to 
adimensionalize the (Coulomb-like) logarithmic divergence contained in the tail.

The last term in Eq. \eqref{UijMPM0} is indicative of a sum of several nonlinear contributions (generated by various couplings between the multipolar waves in the zone exterior to the system). These terms are expressed as products of derivatives of source (and gauge) multipole moments, all taken at the same retarded time $t$ as $U_{ij}(t)$ itself. In $U_{ij}$ at order $G^2/c^5$, there are three such nonlinear contributions: (i) one bilinear in $I_{ij}$ ($U_{ij}^{\rm II}$, denoted $U_{ij}^{\rm QQ}$ in \cite{Bini:2023fiz}); (ii) one bilinear in $I_{ij}$ and the angular momentum $L_k$ ($U_{ij}^{\rm LI} \equiv U_{ij}^{\rm LQ} $), and (iii)  one bilinear in $I_{ij}$ and the $\ell=0$ $W_{i_1\ldots i_\ell}$ gauge moment ($U_{ij}^{\rm WI} \equiv U_{ij}^{\rm WQ} $):
\bea \label{U2nonlin}
U_{ij}^{\rm II + LI + WI}(t)&=&\frac{G}{c^5}\left(\frac17 I_{a\langle i}^{(5)}I^{\vphantom{(5)}}_{j\rangle a}-\frac57 I_{a\langle i}^{(4)}I_{j\rangle a}^{(1)} \right.\nonumber\\
& &\qquad\left.-\;\frac27 I_{a\langle i}^{(3)}I_{j\rangle a}^{(2)} + \ldots\right)\,.
\eea
Though only one power of $G$ enters the prefactor of Eq. \eqref{U2nonlin}, $U_{ij}^{\rm II + LI + WI}(t)$ actually  starts
at the level $G^2$ because of the high-order ($n \geq 3$) time-derivatives of the source quadrupole moment $I_{ij}(t)$
it contains. [As said in Ref.~\cite{Bini:2023fiz} there is a final hereditary term in $U_{ij}(t)$, the nonlinear
memory contribution which, however, starts at order $G^3$, i.e. at the two-loop order.]

When considering the frequency-domain version of $U_2$, we have (at one-loop accuracy, i.e. modulo $O(G^3)$ in $W$)
\be
U_2(\om) = U_2^G(\om)+  U_2^{G^2}(\om)\,,
\ee
where both $U_2^G(\om)$ and $U_2^{G^2}(\om)$ start at Newtonian order. The PN expansions of 
 $U_2^G(\om)$ and $U_2^{G^2}(\om)$ have the structure
\bea
U_2^G &\sim& \frac{GM^2\nu}{p_\infty}\left(1+ \eta^2 p_\infty^2+  \eta^4 p_\infty^4+  \eta^6 p_\infty^6+\ldots\right)\,,\nonumber\\
U_2^{G^2} &\sim&\left(\frac{GM}{bp_\infty^2}\right)\frac{GM^2\nu}{p_\infty} (1+
 \eta^2 p_\infty^2+  \eta^3 p_\infty^3+  \eta^4 p_\infty^4 \nn \\
 && +\;  \eta^5 p_\infty^5+  \eta^6 p_\infty^6+\ldots )\,.
\eea
Note that $\left(\frac{GM}{bp_\infty^2}\right)$ is dimensionless (when considering that $\pinf$ has the
dimension of a velocity)
and of the order of the Newtonian scattering angle.  When considering that $\pinf$ is a velocity, the PN expansion
is encoded in powers of $\eta \,\pinf= \frac{\pinf}{c}$. The tree-level waveform $W^G$ (being time-symmetric)
contains only even powers of $\eta \,\pinf$. By contrast, the one-loop waveform  $W^{G^2}$ starts having
odd powers of  $\eta \,\pinf$ in its fractional PN expansion at order $ \eta^3 p_\infty^3$, which is the order
where hereditary tail effects arise (see Eq. \eqref{U2tail}). 
The next odd power of $\eta \,\pinf$ is $ \eta^5 p_\infty^5$ which
contains several types of contributions coming from: (i) $\eta^5$ and radiation-reaction contributions to Eq. \eqref{U2I};
(ii) fractional 1PN corrections contained in the linear tail term Eq. \eqref{U2tail}; and (iii) the nonlinear contributions
to $U_{ij}$,  Eq. \eqref{U2nonlin}.

Let us finally illustrate the explicit structure of the frequency-domain waveform by displaying a few
important contributions at the $O(G)$ and  $O(G^2)$ levels. 
The Fourier transform is done at the level of the once-time-differentiated,
PN-expanded, Cartesian-component radiative multipoles
$ \frac{d}{dt} U_{ i_1 i_2 \cdots i_{l}}(t)$ and  $\frac{d}{dt} V_{ i_1 i_2 \cdots i_{l}}(t)$. The
 time-dependence of these quantities is conveniently obtained by using the explicit quasi-Keplerian representation
 of the relative two-body motion (in the c.m. frame). The latter representation is
naturally expressed in the $e_x, e_y$ vectorial frame (which is anchored on the classical ${\bar p_a}$ momenta, rather
than the incoming momenta $p_a$). The polar coordinates $r, \varphi$ of the relative motion,
\bea
{\bf x}(t) &=& x(t) e_x + y(t) e_y\nonumber\\
&=& r(t) \left[ \cos \varphi(t) e_x + \sin \varphi(t) e_y \right]\,,
\eea
are given as explicit functions of an (hyperbolictype) ``eccentric anomaly'' variable $v$ by {\it quasi-Keplerian} expressions ons of the form
\bea \label{QK}
{\bar n}\, t &=& e_t \sinh(v) - v + O(\eta^4)\,, \nn\\
r &=& {\bar a}_r \left( e_r \cosh(v) -1) \right) + O(\eta^4)\,,  \\
\varphi &=& 2 K \arctan\left(\sqrt{\frac{e_\varphi+1}{e_\varphi-1}}\tanh\left(\frac{v}{2}\right)\right)   + O(\eta^4)\,.\nn
\eea
Here, the quasi-Keplerian quantities ${\bar n}, e_t, e_r, e_\varphi, K$ are (PN-expanded) functions of the
\com energy and angular momentum of the binary system (see Refs. \cite{DD85,Damour:1988mr,Cho:2018upo}).

The quasi-Keplerian representation Eq.~\eqref{QK} incorporates (in the conservative case) a time symmetry around
$t=0$, corresponding to the closest approach between the two bodies. The asymptotic logarithmic drift of the two
worldlines is embodied in the $v$ parametrization involving hyperbolic functions.

Inserting the (PN-expanded) quasi-Keplerian representation Eq. \eqref{QK} in, say,
\be
- i \om  U_{ i_1 i_2 \cdots i_{l}}(\om)= \int_{-\infty}^{+\infty} dt e^{i \om t} \frac{d}{dt} U_{ i_1 i_2 \cdots i_{l}}(t)\,,
\ee
leads to integrals which can all be obtained (at the $G+G^2$ accuracy) from the master integrals (for $u>0$)
\begin{align} \label{masterK}
\int_{-\infty}^{+\infty} dv e^{i  \left(  u \sinh(v) -  \mu \, v  \right)} &=\int_{-\infty}^{+\infty} \frac{dT}{\sqrt{1+T^2}} 
e^{i    u T -  i \mu \,{\rm arcsinh} T  } \nn \\
&=2e^{\mu\frac{\pi}{2}} K_{ i \mu}(u)\,, \\
\label{masterexp}
\int_{-\infty}^{+\infty} \frac{dv}{\cosh v} e^{i    u \sinh(v) } &=\int_{-\infty}^{+\infty} \frac{dT}{1+T^2} 
e^{i    u T } \nn \\
&=\pi e^{-u}\,.
\end{align}
Actually, one only needs the large-eccentricity expansion (corresponding to a PM expansion in powers of $G$,
and to an expansion in powers of $\mu$) of the master integral Eq. \eqref{masterK}.

Using these integrals, all the SWSH components of the waveform at order $G+G^2$ are expressed 
as linear combinations of  modified Bessel functions of order $0$ and $1$, $K_0(u)$ and $K_1(u)$, and (starting at the 1PN fractional order) of the exponential $e^{-u}/u$, with argument given by
 the following dimensionless version of the frequency:
\be
u \equiv \frac{\om b}{\pinf}\,.
\ee
The coefficients of $K_0(u)$, $K_1(u)$ and $e^{-u}/u$ are ($\nu$-dependent) polynomials in $u$ whose degrees
increase with the PN order.

As already said, when $l=2$ the $m$ values are only $m=+2, m=0$, and $m=-2$. 

\begin{widetext}
Here is a sample of the SWSH components of the quadrupolar piece of the waveform.
At order $G^1\eta^0$  we find
\bea
U_{22}^{G^1\eta^0}&=&  -\frac{GM^2\nu}{p_\infty}\frac{2\sqrt{5\pi}}{5}\Big[(2u+1)K_0(u)+2(u+1)K_1(u)\Big],\nonumber\\
U_{20}^{G^1\eta^0}&=&  -\frac{GM^2\nu}{p_\infty}\frac{2\sqrt{30\pi}}{15}K_0(u),\nonumber\\
U_{2 \bar 2}^{G^1\eta^0}&=&  \frac{GM^2\nu}{p_\infty}\frac{2\sqrt{5\pi}}{5}\Big[(2u-1)K_0(u)-2(u-1)K_1(u)\Big].
\eea
At order  $G^1\eta^2$  we have
\bea
U_{22}^{G^1\eta^2}&=&  \frac{\nu GM^2 p_\infty \sqrt{5\pi}}{105}\Big\{[(-72\nu + 38) u^2 + (-36\nu - 2) u - 24\nu + 78]K_0(u)\nonumber\\
&& +\; [(-72\nu + 38) u^2 + (-72\nu + 17) u - 48\nu - 138]K_1(u)\Big\},\nonumber\\
U_{20}^{G^1\eta^2}&=&  \nu GM^2 p_\infty \frac{\sqrt{30\pi}}{105}\Big[(-8\nu + 26)K_0(u)+u(-16\nu  + 59)K_1(u)\Big],\nonumber\\
U_{2 \bar 2}^{G^1\eta^2}&=&  \frac{\nu GM^2 p_\infty \sqrt{5\pi}}{105}\Big\{[(-72\nu + 38) u^2 + (36\nu + 2) u - 24\nu + 78 ]K_0(u)\nonumber\\
&& + \; [ (72\nu - 38) u^2 + (-72\nu + 17) u + 48\nu + 138]K_1(u)\Big\}.
\eea
The $O(G^2\eta^0)$ multipoles have a simple relation to the  $O(G^1\eta^0)$ ones:
\beq
U_{2m}^{G^2\eta^0}=\frac{\pi}{2}\frac{GM }{bp_\infty^2}\,u\, U_{2m}^{G^1\eta^0}\,.
\eeq
Proceeding to $O(G^2\eta^2)$, we have instead 
\bea
U_{22}^{G^2\eta^2}&=&-\frac{G^2M^3\nu \pi^{3/2}\sqrt{5}}{210 bp_\infty} \Big\{ u  [u^2(72\nu - 38) + u(78\nu  - 124) + 45\nu  - 141]K_0(u)\nonumber\\
&& +\;u [u^2(72\nu  - 38 ) + u(114\nu  - 143) + 90\nu  + 12]K_1(u) +\frac{252(2 u^2 + 2 u + 1)}{u} e^{-u}\Big\}\,, \nonumber\\
U_{20}^{G^2\eta^2}&=& -\frac{G^2M^3\nu \pi^{3/2}\sqrt{30}}{210 bp_\infty} \Big[ u (-47 + 15\nu) K_0(u) + u^2 (-59 + 16\nu) K_1(u)\Big] \,, \nonumber\\
U_{2 \bar 2}^{G^2\eta^2}&=&\frac{G^2M^3\nu \pi^{3/2}\sqrt{5}}{210 bp_\infty} \Big\{ - u [u^2(72\nu   - 38)+u(- 78\nu  + 124)+ 45\nu  - 141 ]K_0(u)\nonumber\\
&& + \; u [u^2(72\nu- 38)+u(  - 114\nu + 143) + 90\nu + 12 ]K_1(u) + \frac{252}{u}e^{-u}\Big\} \,.
\eea
At 2PN the structure of $U_2^{G^2\eta^4}$ is similar to that of $U_2^{G^2\eta^2}$ apart from a quadratic-in-$\nu$ 
dependence of the coefficients and an overall multiplication by  $p_\infty^2$.

Let us finally focus on the $G^2\eta^5$ contribution to $U_2$. Following the decomposition presented above,
each $lm$ component of $U_2$ is the sum of two terms: (i) an ``instantaneous" contribution, $U_{2m}^{\rm inst}$,
defined as 
\beq
U_{2m}^{\rm inst}=U_{2m}^{I}+U_{2m}^{\rm II}+U_{2m}^{\rm LI}+U_{2m}^{\rm WI}\,,
\eeq
i.e. the sum of  the source-moment-related term, $U_{2m}^{I}$,  Eq. \eqref{U2I}, and of the nonlinear
terms indicated in Eq. \eqref{U2nonlin}; and
(ii) a (nonlocal in time) tail term $U_{2m}^{\rm tail}$, Eq. \eqref{U2tail} (here projected on its $G^2 \eta^5$
contribution).
They are respectively given by:
\bea
U_{22}^{{\rm inst}\, G^2\eta^5 }&=&  \frac{2G^2M^3\nu^2\sqrt{5\pi}p_\infty^2}{525 b}i u  \Big[2 (245 u^2 - 57 u - 15) K_0(u) + (490 u^2 + 131 u - 72) K_1(u) \Big]\,,\nonumber\\
U_{20}^{{\rm inst}\, G^2\eta^5 }&=& \frac{2 G^2M^3\nu^2\sqrt{30\pi}p_\infty^2}{105 b}  i u\Big[ -2  K_0(u)+29 u  K_1(u)\Big]\,,\nonumber\\
U_{2\bar 2}^{{\rm inst} \, G^2\eta^5 }&=& \frac{2G^2M^3\nu^2\sqrt{5\pi}p_\infty^2}{525 b} i u  \Big[2 (245 u^2 + 57 u - 15) K_0(u) -  (490 u^2 - 131 u - 72)K_1(u) \Big]\,,
\eea
and
\bea
U_{22}^{{\rm tail}\, G^2\eta^5}&=& -\frac2{105} \frac{G^2M^3\nu\sqrt{5\pi}p_\infty^2}{b} \left(i \log(2 b_0\omega e^{\gamma_E})  - i\frac{11}{12}  + \frac{\pi}{2}\right) u \Big\{[(72\nu - 38) u^2 + (78\nu + 2) u + 45\nu - 78]K_0(u)\nonumber\\
&& +\;[(72\nu - 38) u^2 + (114\nu - 17) u + 90\nu + 138]K_1(u)\Big\}\,,\nonumber\\
U_{20}^{{\rm tail}\, G^2\eta^5}&=&  -\frac2{105} \frac{G^2M^3\nu\sqrt{30\pi}p_\infty^2}{b} \left(i \log(2 b_0\omega e^{\gamma_E})  - i\frac{11}{12}  + \frac{\pi}{2}\right) u \Big[(15\nu - 26)K_0(u)+ u (16\nu - 59)K_1(u)\Big]\,,\nonumber\\
U_{2\bar 2}^{{\rm tail}\, G^2\eta^5}&=& -\frac2{105} \frac{G^2M^3\nu\sqrt{5\pi}p_\infty^2}{b} \left(i \log(2 b_0\omega e^{\gamma_E})  - i\frac{11}{12}  + \frac{\pi}{2}\right) u \Big\{[(72\nu - 38) u^2 + (-78\nu - 2) u + 45\nu - 78]K_0(u)\nonumber\\
&& +\;[(-72\nu + 38) u^2 + (114\nu - 17) u - 90\nu - 138]K_1(u)\Big\}\,. 
\eea
Let us finally give an example of the SWSH components of the $\phi$-odd  (and $(m_1-m_2)$-odd) 
piece of the waveform
\bea \label{V2Geta0}
V_{21}^{G^1\eta^0}&=&-4i \frac{\sqrt{5\pi}}{15}\nu G M^2 \frac{m_1-m_2}{M}u [K_0(u)+K_1(u)]\,,\nonumber\\
V_{2 \bar 1}^{G^1\eta^0}&=&-4i \frac{\sqrt{5\pi}}{15}\nu G M^2 \frac{m_1-m_2}{M}u [K_0(u)-K_1(u)]\,,
\eea
with 
\bea
V_{21}^{G^2\eta^0}&=& \frac{\pi}{2}\left(\frac{GM}{bp_\infty^2}  \right) u \, V_{21}^{G^1\eta^0}\,,\nonumber\\
V_{2 \bar 1}^{G^2\eta^0}&=&\frac{\pi}{2}\left(\frac{GM}{bp_\infty^2}  \right) u \, V_{2\bar 1}^{G^1\eta^0}\,,
\eea
as before.
The corresponding tail contributions  are given by
\bea \label{V2G2eta3}
V_{21}^{G^2\eta^3}&=&  -\frac{4\sqrt{5\pi}}{15b}  \frac{m_1-m_2}{M} \left(\frac73   + \pi i   - 2\log (2b_0 e^{\gamma_E} \omega)\right) p_\infty  G^2M^3\nu u^2 [K_0(u)+K_1(u)]
\nonumber\\ 
&=& -i \frac{GM}{b}p_\infty u  \left( \frac73 i    +i \pi    - 2 i\log (2b_0 e^{\gamma_E} \omega)\right)  V_{21}^{G^1\eta^0},\nonumber\\
V_{2 \bar 1}^{G^2\eta^3}&=&-\frac{4\sqrt{5\pi}}{15b}  \frac{m_1-m_2}{M} \left(\frac73   + \pi i  - 2\log (2b_0 e^{\gamma_E} \omega)\right) p_\infty  G^2M^3\nu u^2 [K_0(u)-K_1(u)]\nonumber\\
&=& -i  \frac{GM}{b}p_\infty u  \left( \frac73     + \pi i    - 2 i\log (2b_0 e^{\gamma_E} \omega)\right) V_{2\bar 1}^{G^1\eta^0} \,.
\eea
All the SWSH components of the waveform $W^{\rm MPM}$
 that we computed are available in the ancillary file  \texttt{MPMmultipoles.nb} attached to this paper.
 They reach (in $W$) $(G^1+G^2) (1 +\eta^2+\eta^3+\eta^4+\eta^5)$
 accuracy for the even $m$ terms, and the   $(G^1+G^2) (\eta + \eta^4)$
 accuracy for the odd $m$ terms.

\end{widetext}

\section{The Observable-based formalism}
\label{sec:KMOC_rev}

Scattering waveform can also be computed using field-theory based methods. In particular, the observable-based formalism~\cite{Kosower:2018adc} relates directly final state observables with $S$-matrix elements in the classical limit. Before we start to introduce the formalism, we note that in field theory, it is customary to define metric perturbation as $g_{\mu\nu} = \eta_{\mu\nu} + \kappa \mathsf{h}_{\mu\nu}$, where $\kappa^2 = 32\pi G$. It differs from $g_{\mu\nu} = \eta_{\mu\nu} + h_{\mu\nu}$ used in the MPM calculation by a factor of $\kappa$. We will stick to the field-theory convention when computing $S$-matrix elements. We will compensate the factor of $\kappa$ when performing the Fourier transform to impact parameter space, see \cref{eq:MtoWEFT} below. We will also set $c=1$ in all the field-theory related calculations.

\subsection{General setup}
\label{sec:setup}

The observable-based formalism \cite{Kosower:2018adc} constructs quantum scattering observables by comparing the expectation value of the corresponding hermitian operator $\opO$ between the initial and final state of the scattering process
\begin{equation}
\langle \opO\rangle=\langle \psi_{\textrm{out}}|\opO|\psi_{\textrm{out}}\rangle-\langle \psi_{\textrm{in}}|\opO|\psi_{\textrm{in}}\rangle \ ,
\label{KMOCgeneral}
\end{equation}
where $\opO$ could be, e.g., the momentum transferred during the scattering process or the curvature at large distances. Observables are assumed to be measured at infinity, where the gravitational field is effectively free. The evaluation of the expectation value in Eq.~\eqref{KMOCgeneral} proceeds by constructing the final state $|\psi_\text{out}\rangle$ as the time-evolution of the initial state $|\psi_\text{in}\rangle$, $|\psi_\text{out}\rangle = \Smatrix |\psi_\text{in}\rangle $,  and then inserting a complete set of final states next to ${\opO}$. The result is an expression of $\langle {\opO}\rangle$ in terms products of $S$-matrix elements suitably integrated over the final-state phase space. 

For the scattering problem, we construct the initial two-particle states as the tensor product of two localized single-particle states,
\begin{align}\label{eq:2pstate}
    |\psi_\text{in}\rangle &= \int \dd\Phi_{p_1}\dd\Phi_{p_2}\varphi(p_1)\varphi(p_2)e^{-i(p_1\cdot b_1+p_2\cdot b_2)}|p_1p_2\rangle\,,\nonumber\\
    \dd\Phi_{p} &= \frac{\dd^4 p}{(2\pi)^4}\Theta(p^0)(2\pi)\delta(p^2+m^2)\,,
\end{align}
where the measure $\dd\Phi_p$ imposes the on-shell condition, and the wavepacket is normalized as $\int\dd\Phi_p|\varphi(p)|^2 = 1$.

To compute the scattering waveform we may  choose the operator ${\opO}$ to either be the linearized Riemann tensor\footnote{The linearization is a consequence of the operator being inserted in the final state, i.e. at infinity. The LSZ reduction then projects out all Green's functions with more than a single graviton.} or to be the transverse-traceless components of the asymptotic metric fluctuations about Minkowski space at infinity.
In the former approach, the metric is extracted from the expectation value of the asymptotic Newman-Penrose scalar 
$\Psi_4\propto {\ddot f}_{+} - i {\ddot f}_{\times}$, by integrating twice over time and determining the two integration constants from physical considerations.
One integration constant is fixed by the boundary condition $\lim_{t\rightarrow \infty}\dot{f}(t) \sim t^{-2}$. The other corresponds to the EFT analog of the Coulombic field in Eq.~\eqref{fin}. 

The latter choice yields directly the waveform. It also gives a matrix-element interpretation of the integration constants needed to reconstruct the asymptotic metric from the Newman-Penrose scalar.
In particular, the integration constant that is time-independent is determined by the parts of Eq.~\eqref{KMOCgeneral} that have support only on vanishing outgoing-graviton frequency $\omega$. Such $\delta(\omega)$-supported matrix elements yield nontrivial time-dependent contributions to the waveform at higher orders in $G$ through iteration, as we will see in~\cref{sec:BMSgeneral}.

The observable-based formalism yields final-state observables in the complete quantum theory. 
The building block of the momentum-space expectation value of the classical asymptotic metric is
\begin{align}\label{eq:KMOC_ME}
    \MKMOC & (\varepsilon,k,p_1,p_2,q_1,q_2)\nonumber\\
    &\equiv i \langle p_3 p_4 | \opa(k) \opT  |p_1 p_2\rangle + \langle p_3 p_4|\opT^{\dagger}\opa(k)\opT|p_1 p_2\rangle\nonumber\\
    &=  i \langle p_3 p_4 k| \opT  |p_1 p_2\rangle + \langle p_3 p_4|\opT^{\dagger}\opa(k)\opT|p_1 p_2\rangle \,,
\end{align}
where $\opa$ creates a physical graviton with polarization vector $\varepsilon$ when acting to the left.
The momentum lost by each matter particle is
\begin{align}
    q_1 = p_1 - p_4\,,\qquad q_2 = p_2 - p_3\,,
\end{align}
and momentum conservation requires that $q_1+q_2$ be the outgoing graviton momentum $k$.
The first term in $\MKMOC$ is the five-point amplitude with one graviton in the final state; the second term is a unitarity cut that gives a particular discontinuity of the quantum amplitude. We shall refer to this bilinear-in-$T$ term as the ``unitarity cut term'', or simply, the ``cut term''. 
The matrix element $\MKMOC$ is a Lorentz invariant function of the various scalar products of $(\varepsilon,k,p_1,p_2,q_1-q_2)$. For notational convenience, in the following we will (partially) suppress the arguments of~$\MKMOC$.

The absence of local diffeomorphism-invariant observables in general relativity requires that one be careful with the effect of large gauge transformations on asymptotic observables. The asymptotic Newman-Penrose scalar and the transverse-traceless components of the asymptotic metric are invariant under decaying-at-infinity coordinate transformations, $\delta h_{\mu\nu} = \partial_{(\mu} \xi_{\nu)}$, but are not invariant under BMS transformations. However, as we will see in Section \ref{sec:BMSgeneral}, both the KMOC and MPM formalisms (implicitly) use the same BMS vacuum, referred to as ``intrinsic gauge'' in Ref. \cite{Veneziano:2022zwh}.\footnote{
Perhaps the term ``intrinsic frame'' might be more appropriate as it should be possible to discern its properties through gravitational wave measurements.
}

\subsection{Classical limit}

As we reviewed above, the observable-based formalism establishes the relation between observables with complete velocity dependence and the matrix element $\MKMOC$ in the quantum theory. The correspondence principle, stating that the classical regime emerges in the limit of large (macroscopic) conserved charges,  provides a means to extract the classical observables from their quantum mechanical counterpart. In our case these charges are the particles' masses, momenta and orbital angular momenta $J$, i.e. $J\gg \hbar = 1$ and $m\gg M_{\rm Pl}$. It then follows \cite{Cheung:2018wkq, Kosower:2018adc, Bern:2019crd} that the impact parameter $b$ is much larger than the Compton wave length $\lambda_c = p^{-1}$ or, equivalently, that the momentum transfer $q$ is much smaller than the typical external momenta $p$ \footnote{
In later sections we will compare MPM and amplitudes-based predictions in the PN expansion. To be in the regime of validity of both this and of the PM expansions originally used in the amplitudes computation, we must have the stronger condition on the impact parameter, $b\gg v^{-2} R_s$, corresponding to a small scattering angle $\chi \sim GM/( b v^2)$.} and that observables are series in the effective coupling $G m_1 m_2/J$.

In this limit, certain parts of loop-level expressions contribute to classical  observables. The precise dependence on Newton's constant $G$ and momentum transfer 
follows from the observation that, like Newton's potential itself, powers of Newton's potential are also classical; in impact parameter space, a classical $L$-loop contribution is proportional $(G/b)^{L+1}$. The inclusion of outgoing (gravitational) radiation with energy $\omega$ commensurate with the momentum transfer, $\omega\sim q$, does not alter this dependence because there is no suppression for the emission of classical (gravitational) radiation. 
Quantum scattering amplitudes are however more singular in the large impact parameter (small transferred momentum) expansion than their classical counterparts, behaving $G^{L+1}/b^{l<L+1}$, with additional mass-dependent factors. This can be understood by recalling the structure of potential scattering in non-relativistic quantum mechanics, which at $N$-th order in perturbation theory contains multiple insertions of the lower-order potential. The cancellation of these classically singular (or superclassical) terms in the matrix element $\MKMOC$ is a test of the calculation.

It is extremely convenient to take the classical limit as early as possible~\cite{Cheung:2018wkq, Bern:2019nnu, Bern:2019crd}, \textit{before} the Feynman loop integrals are evaluated, as this substantially reduces the complexity of calculations.  Each of the internal graviton exchanged between matter particles should mediate a long-range interaction; thus, in this regime all loop momenta $\ell$ should be of the same order as the total exchanged momentum $q$. 
The method of regions~\cite{Beneke:1997zp} in dimensional regularization is a systematic method to select the classical contributions to $\MKMOC$. The 
integration domain is split into two regions
\begin{equation}\label{eq:regions}
\begin{split}
\textrm{Hard Region}: \quad (q_i,k,\ell)&\rightarrow(\lambda q_i,\lambda k,\lambda^{0}\ell) \ , \\
\textrm{Soft Region}: \quad (q_i,k,\ell)&\rightarrow(\lambda q_i,\lambda k,\lambda\ell) \ ,
\end{split}
\end{equation}
and the latter generates all classical terms. To evaluate an integral in only one of the regions one simply expands the integrand according to the indicated scaling in $\lambda$ and integrates over the {\em entire} integration domain. The full integral is given by the sum of the integrals in the two regions~\cite{Beneke:1997zp}; there is no overcounting because the integrand in one region yields only scaleless integrals when further expanded in the other region, and such integrals vanish in dimensional regularization.

The momentum space KMOC matrix element $\MKMOC$ is related to the frequency-domain EFT waveform $\WEFT$ (here normalized as the MPM waveform $W^{\rm MPM}$) through a Fourier transform to impact-parameter space, 
\begin{align}\label{eq:MtoWEFT}
    &\WEFT  (\varepsilon,k,p_1,p_2,b_1,b_2) \\
    & = \frac{2i}{\kappa} \int \mu(k,q_1,q_2) e^{-i(q_1\cdot b_1+q_2\cdot b_2)}\MKMOC(\varepsilon,k,p_1,p_2,q_1,q_2)\,,\nonumber
\end{align}
and we choose the (incoming) impact parameter $b_{12} = b_1-b_2$ to be orthogonal to both $p_1$ and $p_2$, 
\begin{equation}
    p_1\cdot b_{12}= p_2\cdot b_{12}=0\,.
\end{equation}
The integration measure enforces momentum conservation and on-shell conditions for the matter particles. Following \cref{eq:2pstate}, we can write the measure as 
\begin{align}
    &\int\dd\Phi_{p_1}\dd\Phi_{p_2}\dd\Phi_{p_3}\dd\Phi_{p_4}\varphi^{*}(p_3)\varphi^{*}(p_4)\varphi(p_1)\varphi(p_2)\nonumber\\
    &\hspace{3.35cm}\times \hdelta^4(p_1+p_2-p_3-p_4-k)\nonumber\\
    &=\int \mu(k,q_1,q_2)  + (\text{quantum})\,,
\end{align}
and the classical contribution is
\begin{align}\label{eq:cl_measure}
    \mu(k,q_1,q_2) = \frac{\dd^4 q_1}{(2\pi)^4}\frac{\dd^4 q_2}{(2\pi)^4} & \hdelta(2p_1\cdot q_1)\hdelta(2p_2\cdot q_2)\\
    &\times\hdelta^{4}(q_1+q_2-k)\,, \nonumber
\end{align}
where $\hdelta(x) = 2\pi\delta (x)$ and $\hdelta^{d}(x) = (2\pi)^d \delta^d(x)$.
We now sketch the derivation of \cref{eq:cl_measure}.
In the classical limit, the Compton wavelength $\ell_{\text{c}}$ is much smaller than the spread of the wavepacket $\ell_{\phi}$ in the position space, which is in turn much smaller than the impact parameter $|b_1-b_2|$, namely, $\ell_{\text{c}}\ll\ell_{\phi}\ll|b_1-b_2|$. 
Consequently, in a classical scattering process, the wavepackets $\varphi(p_1)$ and $\varphi^{*}(p_4)$ exactly overlap up to quantum corrections, which allows us to integrate out the wavepackets, $\int\dd\Phi_{p_1}\varphi(p_1)\varphi^{*}(p_4)\simeq 1$ [same for $\varphi(p_2)$ and $\varphi^{*}(p_3)$]. Then in $\dd\Phi_{p_3}$ and $\dd\Phi_{p_4}$ we change the variables to $q_1$ and $q_2$ while keeping only the leading term in the on-shell condition,
\begin{align}
    \delta(p_4^2 + m_1^2) &\simeq \delta (2 p_1\cdot q_1)\,,\nonumber\\
    \delta(p_3^2 + m_2^2) &\simeq \delta (2 p_2\cdot q_2)\,.
\end{align}
Note that we can drop the positive energy constraint in the measure because it is already implicit in the above delta function constraints.

\subsection{IR divergence and time shift}
\label{sec:IR}

The matrix element $\MKMOC$ is infrared-divergent \cite{Herderschee:2023fxh, Elkhidir:2023dco, Brandhuber:2023hhy, Georgoudis:2023lgf, Caron-Huot:2023vxl}, with contributions from both terms in Eq.~\eqref{eq:KMOC_ME}, which we regularize in dimensional regularization with $d=4-2\epsilon$.  Even at the one-loop order we are focusing on, the finite waveform integrand depends on whether internal states are four- or $d$-dimensional. This is in contrast with scattering angle 
calculations, where this subtlety appears only at $O(G^4)$~\cite{Bern:2021yeh}, and is due here to the presence of divergences which are effectively (additively) renormalized. 
To avoid possible issues probably due to unconventional forms of dimensional regularization (see e.g. \cite{Harlander:2006rj,Harlander:2006xq}), we use conventional dimensional regularization (CDR)~\cite{Collins:1984xc}, where all states and momenta are uniformly continued to $d=4-2\epsilon$ dimensions. 
In the final expressions we restrict the external momenta to be four dimensional and use special four-dimensional relations such as vanishing Gram determinants for further simplifications. 
This regularization scheme was important in the comparison of the ${O}(G^4)$ scattering angle~\cite{Bern:2021yeh} and the corresponding post-Newtonian calculations of Ref.~\cite{Bini:2021gat}; this will be the case here as well.

The IR divergences in $\MKMOC$ are due to virtual soft gravitons and are expected to exponentiate~\cite{Weinberg:1965nx},
\begin{align}\label{eq:IR_exp}
    \MKMOC = \exp\left[-i \omega \frac{GE}{\epsilon} ( 1 -\Gamma/2)\right] \MKMOC^{\text{fin}}\, , 
\end{align}
where $\MKMOC^{\text{fin}}$ is IR-finite but dependent of $\epsilon$. 
The first term in the exponent originates from the Weinberg soft factor from the five-point amplitude~\cite{Weinberg:1965nx}, whose proof can be 
generalized to all orders in the dimensional regulator.
The second term comes from the  cut term~\cite{Caron-Huot:2023vxl} and $\Gamma$ is defined as
\begin{align}
    \Gamma = \frac{3\y-2\y^3}{(\y^2-1)^{3/2}} \ .
\end{align}
The factor  $\omega E$ in the exponent of \cref{eq:IR_exp} can be written covariantly as 
\begin{align}
\label{wToE}
    m_1 w_1 + m_2 w_2 = \omega E\,,
\end{align}
where $w_1$ and $w_2$ are
\begin{align}
    w_a = -u_a\cdot k = -\frac{p_a\cdot k}{m_a}\,.
\end{align}

To understand the fate of the IR divergence, it is useful to consider the time-domain waveform, which is related to the frequency-domain waveform~\eqref{eq:MtoWEFT} by the Fourier-transform
\begin{align}
    \WEFT(t) = \int_{-\infty}^{+\infty}\frac{\dd\omega}{2\pi} \, \WEFT(\omega) \, e^{-i\omega t}\,.
\end{align}
The IR divergent phase can now be absorbed in the (re)definition of the retarded time,
\begin{align}
    t\rightarrow t+\delta t_{\text{IR}}\,, \quad \delta t_{\text{IR}} = \frac{GE}{\epsilon}(1-\Gamma/2) \,.
    \label{timeshift_IR}
\end{align}
Equivalently, we can choose to use the IR finite matrix element in \cref{eq:MtoWEFT} when computing the waveform,
\begin{align}\label{eq:MtoMfin}
    \MKMOC \rightarrow \MKMOC^{\text{fin}} = \exp\left[i \omega \frac{GE}{\epsilon} ( 1 - \Gamma/2 )\right] \MKMOC \,.
\end{align}
The result for this IR-finite matrix element at $L$ loops depends on the ${O}(\epsilon^{1\le n\le L})$ part of $\MKMOC$, which in 
turn depends on the dimensionality of the internal states at lower-loop orders.  
In particular, the tree-level term in $\MKMOC$ contains such ${O}(\epsilon)$ contributions which enforce that the regularized theory 
contains only $d$-dimensional graviton states, which we will refer to as $\epsilon \mathcal{M}^{(0)}_{\text{extra}}$. While they are unimportant 
for the  $O(G)$ waveform, they interfere with the $O(G/\epsilon)$ argument of the phase and yield finite contributions 
at $O(G^2)$ proportional to $\mathcal{M}^{(0)}_{\text{extra}}$. 
At one-loop order inclusion of these terms amounts to a choice of scheme, and the one used here is different from that 
used in~\cite{Herderschee:2023fxh, Elkhidir:2023dco, Brandhuber:2023hhy, Georgoudis:2023lgf, Caron-Huot:2023vxl}. 
Their inclusion however is natural in light of the fact that Weinberg's exponentiation of IR divergences holds to all 
orders in dimensional regulator in CDR suggesting that they are required for the exponentiation at two and higher loops. 
Furthermore, this scheme led to a successful comparison of amplitudes and PN predictions for the ${O}(G^4)$ conservative scattering angle, where similar $\epsilon/\epsilon$ effects played an important role.  

\section{Waveform from observable-based formalism}
\label{sec:waveform_KMOC}

Based on their origin in the $T$-matrix element, we can divide momentum-space matrix element $\MKMOC$ into three categories,
\begin{align}
\label{eq:MKMOC_class}
    \MKMOC(k,q_1,q_2) &= \MKMOC_{\text{const}}(k,q_1,q_2) + \MKMOC_{\text{conn}}(k,q_1,q_2) \nonumber\\
    &\quad + \MKMOC_{\text{disc}}(k,q_1,q_2) \,.
\end{align}
The first term $\MKMOC_{\text{const}}$ is contributed by matrix elements that have the external graviton attached to an on-shell three-point vertex. Due to momentum conservation, the graviton must have zero energy. Thus $\MKMOC_{\text{const}}$ corresponds to a constant gravitational background. The second term $\MKMOC_{\text{conn}}$ contains the amplitudes and unitarity cuts contributed from connected $T$-matrix elements with generic momentum, while $\MKMOC_{\text{disc}}$ contains the unitarity cuts containing disconnected $T$-matrix elements.
We note that both $\MKMOC_{\text{const}}$ and $\MKMOC_{\text{disc}}$ receive contributions from the disconnected $T$-matrix elements, as shown in \cref{fig:disconnected}. We discuss in turn $\MKMOC_{\text{const}}$, $\MKMOC_{\text{conn}}$ and $\MKMOC_{\text{disc}}$ in 
\cref{sec:constant_background,ampANDconnectedcuts,sec:BMSgeneral} below. Note that $\MKMOC_{\text{const}}$ is singled out by being constant in retarded time.

In this section, the matter momentum $p_a$ and four-velocity $u_a$ are to be understood as the ``barred variables'' in amplitude literature (see e.g.~\cite{Luna:2017dtq, Parra-Martinez:2020dzs}). They differ from the true momentum $\mathsf{p}_{a}$ (used e.g. in \cref{sec:background2}) by half of the momentum transfer $q_a$, i.e. $p_a = \mathsf{p}_a + q_a/2$. In the classical limit, the difference between $p_a$ and $\mathsf{p}_a$ is quantum so they are indistinguishable. In contrary, the $\pb_a$ used in the MPM calculation is the ``(classical) averaged momentum'', defined as $\pb_a = p_a + \Delta p_a/2$, where $\Delta p_a$ is the classical impulse.

\subsection{Constant background}
\label{sec:constant_background}

Exactly-zero energy excitations are an interesting feature of quantum field theories with massless particles. On the one hand, the inclusion of an arbitrary number of such excitations does not change the energy of any state, such as the Higgs field in spontaneously-broken gauge theories~\cite{Craig:2011ws, Kiermaier:2011cr}. On the other, they, and their off-shell extensions, may condense into nontrivial time-independent backgrounds, such as the Coulomb field of a point charge or the Schwarzschild metric~\cite{Duff:1973zz, Bjerrum-Bohr:2018xdl}. 
From the perspective of on-shell scattering amplitudes around flat space, only the asymptotic zero-energy states appear to be relevant. It has been suggested in Ref.~\cite{Cristofoli:2022phh} that their effects may be incorporated in a vacuum with no such excitations through a coherent state.

It was emphasized in Ref.~\cite{Veneziano:2022zwh} that the constant part of the waveform is dependent on the choice of BMS supertranslation frame. 
The so-called ``canonical frame'' is defined to have a vanishing time-independent component and the ``intrinsic frame'' having the non-vanishing value given in Eq.~\eqref{fin}.
A BMS supertranslation connects the waveforms in the two frames; in light of Ref.~\cite{Cristofoli:2022phh} we may expect that this supertranslation is captured by coherent state of low/vanishing-energy gravitons.
Instead of exploring the all-order nature of this expectation, we will study the effect of including zero-energy gravitons in the KMOC computation.

\begin{figure}[t]
    \centering
    \subfloat[]{\label{disconnected5ptTree}\begin{tikzpicture}
        \pgfmathsetmacro{\sc}{0.7};        
        \path (0,0) pic [scale=\sc] {discTree};
    \end{tikzpicture}}\qquad
    \subfloat[]{\label{disconnected5ptCut}\begin{tikzpicture}
        \pgfmathsetmacro{\sc}{0.7};    
        \path (0,0) pic [scale=\sc] {discOneLoopBG};
    \end{tikzpicture}}
    \caption{The contribution to the KMOC matrix element from the disconnected components of the $T$-matrix. The first diagram contributes to $\MKMOC_{\text{const}}$ and the second one contributes to $\MKMOC_{\text{disc}}$. Note that one of the $T$ matrices on the right of the cut in the second diagram is disconnected. See the discussion below \cref{eq:MKMOC_class} for their classification.} 
    \label{fig:disconnected}
\end{figure}
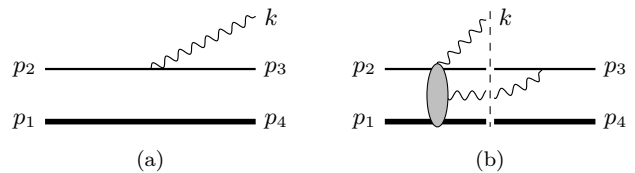

The leading order contribution to $\MKMOC_{\text{const}}$ comes from the diagram shown in \cref{disconnected5ptTree} together with its $(p_1 p_4) \to (p_2 p_3)$ counterpart, i.e. one matter line does not emit any graviton while the other one emits a zero-energy (and therefore zero-momentum) on-shell graviton. 
The relevant amplitudes are
\begin{align} 
    \mathcal{M}_3(k,p_1,p_4) &= - i \kappa (\varepsilon\cdot p_1)^2\,, \nonumber\\
    \mathcal{M}_3(k,p_2,p_3) &= - i \kappa (\varepsilon\cdot p_2)^2\,.
\end{align}
For on-shell and real momenta, this process is supported at the origin of phase space, at which the emitted graviton has exactly zero energy. 

Since one matter particle does not participate in the scattering, we need to use the three-particle phase space to perform the Fourier transform,
\begin{align} \label{WEFTin}
    \WEFT_{\text{const}}(k)\Big|_{m_1} &= \frac{2i}{\kappa}\int\dd\Phi_{p_4}\hdelta^4(p_1-p_4-k)e^{-i(p_1-p_4)\cdot b_1}\nonumber\\
    &\hspace{2cm}\times \mathcal{M}_3(k,p_1,p_4) \nonumber\\
    &= m_1 (\varepsilon\cdot u_1)^2 \hdelta(u_1\cdot k)\, .
\end{align}
%
The delta function in Eq. \eqref{WEFTin} has a unique solution for which the on-shell graviton has exactly zero momentum. 

Thus, if exacty zero energy gravitons are excluded from the spectrum of states, $\WEFT_{\text{const}}(k) = 0$ and we recover the expected result in the BMS canonical frame. Consistently excluding such gravitons from higher-order calculations amounts to excluding all kinematic configurations which constrain at least one graviton to have exactly zero energy.   

Conversely, allowing such contributions reproduces the MPM Coulombic background given in \cref{fin,Win}.\footnote{We note that the ``constant background'' here is different from $\mathcal{M}^{\text{lin}}(k)$ given in Eq.~(5.2) of \cite{Bini:2023fiz}. Here the background is chosen to be the pure time-independent piece of the waveform that is present at $t=-\infty$, while $\mathcal{M}^{\text{lin}}(k)$ there was defined as the average of the time-independent contributions at $t=\pm\infty$. In other words, $\mathcal{M}^{\text{lin}}(k)$ included some time-dependent evolutions that contribute to the gravitational-wave memory.} \footnote{The observation that zero-energy gravitons reproduce the Coulombic background was also made recently in a related context in~\cite{Adamo:2024oxy}.} At higher orders, it requires us to consistently include special kinematic configurations localizing at least one graviton at the origin of phase space. Furthermore, we may expect that these terms may be obtained from those supported on finite-energy configurations through the same supertranslation as the one connecting the intrinsic and canonical frames. 

As we have discussed in \cref{sec:setup}, the Coulombic background serves to fix the initial value of the waveform. Since the three-point amplitudes are sufficient to match the initial condition with the MPM calculation, other diagrams involving a zero energy graviton are not expected to contribute to the constant background. 
We will show in \cref{sec:background2} that it is indeed the case.

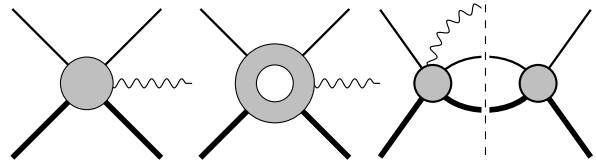
\begin{figure}[t]
    \centering
    \begin{tikzpicture}
        \pgfmathsetmacro{\sc}{0.7};        
        \path (0,0) pic [scale=\sc] {treeAmp};
        \path (2.5,0) pic [scale=\sc] {oneLoopAmp};
        \path (5.3,0) pic [scale=\sc] {oneLoopCut};
    \end{tikzpicture}
    \caption{The connected contribution to the KMOC matrix element. Each blob here is a connected component of the $T$-matrix. The first two graphs are respectively the tree and one-loop amplitude, and the last graph is a unitarity cut. The external particles are labeled in the same way as \cref{fig:disconnected}.}
    \label{fig:connected}
\end{figure}

\subsection{Amplitudes and connected cuts}
\label{ampANDconnectedcuts}

The contribution of all the connected (generic momentum) $T$-matrix elements to $\MKMOC$ have been evaluated in~\cite{Herderschee:2023fxh, Elkhidir:2023dco, Brandhuber:2023hhy, Georgoudis:2023lgf} to $O(\kappa^5)$. We can write it as
\begin{align}
    \mathscr{M}_{\text{conn}}(k,q_1,q_2) &= 
    \mathcal{M}^{\text{tree}} + \mathcal{M}^{\text{1 loop}} + \mathcal{S}^{\text{1 loop}} + O(\kappa^7)\,,
\end{align}
where $\mathcal{M}^{\text{tree}}$ and $\mathcal{M}^{\text{1 loop}}$ are respectively the tree-level and one-loop amplitude contributed from the first term (linear in $T$) of \cref{eq:KMOC_ME}, while the cut contribution $\mathcal{S}^{\text{1 loop}}$ comes from the second term of \cref{eq:KMOC_ME}. Diagrammatically they are represented by \cref{fig:connected}.

For later convenience, we separate out the $\kappa$ and mass dependence from $\mathcal{M}^{\text{tree}}$,
\begin{align}\label{eq:5ptTreeFull}
    \mathcal{M}^{\text{tree}} = \kappa^3 m_1^2 m_2^2 \mathcal{M}^{(0)}_{d}\,.
\end{align}
The reduced tree-level amplitude is given by
\begin{eqnarray} \label{eq:5ptTree}
 \mathcal{M}^{(0)}_{d}&=& i \left[ -F_2^2\frac{y^2}{q_1^2q_2^2 w_2^2} \right. \\
 &+&F_3^2 \left(-\frac{1}{q_1^2 q_2^2} +\frac{y}{q_2^2w_1w_2}+\frac{y^2}{4w_1^2w_2^2}-\frac{q_1^2 y^2}{4q_2^2 w_1^2w_2^2}\right)\nonumber\\
 &+&\left. F_2F_3 \left(-\frac{2y}{q_1^2 q_2^2w_2}+\frac{y^2}{q_2^2 w_1 w_2^2}\right)\right] \nonumber\\
 &+& \frac{i}{d-2} \left[\frac{F_3^2(q_1^2-q_2^2)}{4 q_2^2 w_1^2w_2^2 } 
 +  \frac{F_2^2}{q_1^2q_2^2w_2^2}  - \frac{F_2 F_3}{q_2^2 w_1 w_2^2}\right].\nonumber
\end{eqnarray}
where the graviton polarizations are encoded in the gauge invariant variables
\begin{align}
    F_1 = F_{\mu\nu}q_1^{\mu}u_1^{\nu}\,, \quad F_2 = F_{\mu\nu} q_2^{\mu} u_2^{\nu}\,,\quad F_3 = F_{\mu\nu} u_1^{\mu} u_2^{\nu}\,,    
\end{align}
and $F_{\mu\nu}=k_{\mu}\varepsilon_{\nu} - k_{\nu}\varepsilon_{\mu}$ is the linearized field strength. For $d=4-2\epsilon$, we have, 
\begin{align}\label{eq:Mtree}
    \mathcal{M}_{d=4-2\epsilon}^{(0)} = \mathcal{M}^{(0)}_{d=4} + \epsilon \mathcal{M}^{(0)}_{\text{extra}} + O(\epsilon^2)\,,
\end{align}
where the extra dimensional contribution comes from the $\frac{1}{d-2}$ factor in \cref{eq:Mtree},
\begin{align} \label{eq:M0extra}
{\mathcal M}^{(0)}_{\rm extra}
= \frac{i[4 F_2^2w_1^2 - 4 F_2 F_3w_1q_1^2 + F_3^2 q_1^2 (q_1^2-q_2^2)]}{8(q_1^2q_2^2 w_2^2w_1^2)}\,. 
\end{align}
Although the external states are kept in exactly four dimensions, the $\epsilon$ dependence appears here to ensure that only graviton states propagate in $d=4-2 \epsilon$ dimensions.

This is a consequence of the dimensional regularization scheme we choose.
Of course, we can set $d=4$ when computing the tree-level waveform, and $\mathcal{M}^{(0)}_{\text{extra}}$ does not contribute.

Similarly, we can write the one-loop amplitude and cut contribution as
\begin{align}\label{eq:M1L}
    \mathcal{M}^{\text{one loop}} &= \kappa^5 m_1^3 m_2^2 \mathcal{M}_{1}^{(1)} + \kappa^5 m_1^2 m_2^3 \mathcal{M}_{2}^{(1)}\,,\nonumber\\
    \mathcal{S}^{\text{one loop}} &= \kappa^5 m_1^3 m_2^2 \mathcal{S}_{1}^{(1)} + \kappa^5 m_1^2 m_2^3 \mathcal{S}_{2}^{(1)}\,.
\end{align}
The reduced one-loop amplitude $\mathcal{M}_{1}^{(1)}$ is given by
\begin{align}\label{eq:amp1L}
    \mathcal{M}_{1}^{(1)} = - \frac{i w_1}{32\pi\epsilon} \mathcal{M}^{(0)}_{d=4} + \widetilde{\mathcal{M}}_{1}^{(1)}\,,
\end{align}
where the IR finite part $\widetilde{\mathcal{M}}_{1}^{(1)}$ is the same as Eq.~(5.6) of~\cite{Bini:2023fiz}. The expression of $\widetilde{\mathcal{M}}_{1}^{(1)}$ can be found in \cref{app:amp}.
We note that $\mathcal{M}_{2}^{(1)}$ in \cref{eq:M1L} can be obtained from $\mathcal{M}_{1}^{(1)}$ by exchanging the particle labels $1$ and $2$. On the other hand, the cut contribution can be written as
\begin{align}\label{eq:cut1L}
    \mathcal{S}^{(1)}_{1} &= \frac{iw_1\Gamma}{64\pi\epsilon}\mathcal{M}^{(0)}_{d=4} + \widetilde{\mathcal{S}}_1^{(1)}\,,
\end{align}
where the finite part $\widetilde{\mathcal{S}}_1^{(1)}$ is given in \cref{app:amp}, and $\mathcal{S}_{2}^{(1)}$ is given by the same particle relabeling. Note that we have removed a local UV divergence from $\mathcal{S}_{1}^{(1)}$. Such terms do not affect the long-range interactions that we are interested in, and they are projected out by the Fourier transform to impact parameter space.

The full IR divergence of the one-loop waveform is
\begin{align}\label{eq:1LIR}
    (\mathcal{M}^{\text{1 loop}} + \mathcal{S}^{\text{1 loop}})\Big|_{\text{IR}} = - \frac{i \omega \kappa^2 E(2-\Gamma)}{64\pi\epsilon}\mathcal{M}^{\text{tree}}_{d=4}\,,
\end{align}
Following \cref{eq:MtoMfin} and the discussion below, we can absorb the IR divergence into an infinite shift in the retarded time $\delta t_{\text{IR}} = \frac{G E}{\epsilon} (1-\Gamma/2)$ to obtain the IR finite momentum-space waveform,
\begin{align}
    \MKMOC^{\text{fin}} &= \exp\left[i \omega \frac{GE}{\epsilon} (1-\Gamma/2)\right]\MKMOC \nonumber \\
    &= \mathcal{M}^{\text{tree}} + \mathcal{M}^{\text{1 loop}} + \mathcal{S}^{\text{1 loop}} \nonumber\\
    &\quad + \frac{i \omega \kappa^2 E(2-\Gamma)}{64\pi\epsilon}\mathcal{M}^{\text{tree}} + O(\kappa^7)\\
    &= \mathcal{M}^{\text{tree}} + \mathcal{M}^{\text{1 loop fin}} + \mathcal{S}^{\text{1 loop fin}} + O(\kappa^7)\,.\nonumber
\end{align}
Crucially, the time shift introduces a term proportional to the $d$-dimensional tree amplitude $\mathcal{M}^{\text{tree}}$. While this term cancels the IR divergence~\eqref{eq:1LIR}, it brings an additional finite contribution proportional to the extra dimensional component~\eqref{eq:M0extra}.
Organizing the finite one-loop amplitude and cut by the mass dependence,
\begin{align}\label{eq:MCfiniteFull}
    \mathcal{M}^{\text{1 loop fin}} &= \kappa^3 m_1^3 m_2^2 \mathcal{M}_1^{(1)\text{ fin}} + \kappa^3 m_1^2 m_2^3 \mathcal{M}_2^{(1)\text{ fin}}\,,\nonumber\\
    \mathcal{S}^{\text{1 loop fin}} &= \kappa^3 m_1^3 m_2^2 \mathcal{S}_1^{(1)\text{ fin}} + \kappa^3 m_1^2 m_2^3 \mathcal{S}_2^{(1)\text{ fin}}\,,
\end{align}
we get
\begin{align}\label{eq:MCfinite}
    \mathcal{M}_{1}^{(1)\text{ fin}} &= \widetilde{\mathcal{M}}_{1}^{(1)} + \frac{i w_1}{32\pi}\mathcal{M}^{(0)}_{\text{extra}}\,,\nonumber\\
    \mathcal{S}_{1}^{(1)\text{ fin}} &= \widetilde{\mathcal{S}}_{1}^{(1)} - \frac{i w_1 \Gamma}{64\pi} \mathcal{M}^{(0)}_{\text{extra}} \,.
\end{align}
Comparing with \cref{eq:amp1L,eq:cut1L} we see that, while the divergence is removed by absorbing the $1/\epsilon$ into a time shift, this shift introduces a new contribution proportional to the extra dimensional component of the five-point tree amplitude defined in \cref{eq:Mtree}. Thus the additional terms proportional to $\mathcal{M}^{(0)}_{\text{extra}}$ come from the $\epsilon/\epsilon$ cancellation, and they are the direct consequence of the regularization scheme chosen in \cref{sec:IR}. While at this loop order the inclusion of this term may be regarded as a scheme dependence (and was not included in previous amplitude calculations or comparisons), its presence here is justified by the expected higher-loop structure of IR divergences as well as by its importance in related contexts~\cite{Bern:2021yeh}.
As we will see in Sec.~\ref{sec:Comparison}, its presence substantially improves the comparison between MPM and amplitudes-based results.

\subsection{Cuts with disconnected matrix elements and BMS transformation}
\label{sec:BMSgeneral}

If zero-energy gravitons are excluded, then all cut contributions containing a three-point S-matrix element or an $n+1$-point matrix element in the $1\rightarrow n$ configuration vanish identically. All disconnected contributions to the cut term in Eq.~\eqref{eq:KMOC_ME} are of this type.  
If zero-energy gravitons are kept in the spectrum, the (albeit degenerate) three-point amplitude generates, at higher-order in Newton's constant, time-dependent terms  through the cut term in this equation.
 The relevant contributions at one-loop order are illustrated in Fig.~\ref{disconnected5ptCut}; the discussion above implies that, at $L$ loops, the six-point tree amplitude factor is replaced by the corresponding $(L-1)$-loop six-point amplitude with no changes to the disconnected factor.\footnote{In the full quantum theory the three-point Green's function is UV divergent; this divergence arises from loop momenta outside the soft region, and do not contribute to the classical observables we are discussing here.}

The evaluation of this cut makes use of the fact that the cut massless momentum $\ell$ must have vanishing frequency, so only the leading soft limit of the six-point tree amplitude (the left blob of \cref{disconnected5ptCut}) with two external gravitons of momenta $k$ and $\ell$ is relevant,
\begin{align}
\label{eq:6ptSoft}
{\cal M}_6(p_{1\ldots4}, k, \ell) &\simeq -\kappa\,\mathcal{M}_5(p_{1\ldots4}, k)
\\
&\quad \times \left(
\sum_{i=a}^4 \frac{\eta_a\varepsilon(\ell)_{\mu\nu}p_a^\mu p_a^\nu}{2\ell\cdot p_a-\ie\eta_a}
+\frac{\varepsilon(\ell)_{\mu\nu}k^\mu k^\nu}{2\ell\cdot k-\ie}
\right),
\nonumber
\end{align}
where $\eta_a=\pm 1$ for outgoing and incoming particles, respectively, and $\mathcal{M}_5(p_{1\ldots4},k)$ is the five-point amplitude with one external graviton of momentum $k$.

Sewing \cref{eq:6ptSoft} onto the disconnected five-point amplitude, we find that the four summed terms in \cref{eq:6ptSoft} that involve massive matter momenta will lead to integrals of the form
\begin{align}\label{eq:zero_int}
    &\int\dd^d \ell \,\hdelta(u_3\cdot\ell)\hdelta(\ell^2)\Theta(\ell^0)\,,\nonumber\\
    &\int\dd^d \ell \,\hdelta(u_3\cdot\ell)\hdelta(\ell^2)\hdelta(u_4\cdot\ell) \Theta(\ell^0)\, ,
\end{align}
both of which are integrated to zero in $d=4-2\epsilon$.
Therefore, a nonzero contribution can only come from the last term in this equation, which corresponds to the soft graviton being attached to the outgoing graviton with momentum $k$. We can replace $\mathcal{M}_{5}(p_{1...4},k)$ by its classical limit given in \cref{eq:5ptTreeFull}. It then yields
the following contribution to $\MKMOC_{\text{disc}}$ at ${O}(\kappa^5)$ (which makes use of the on-shell condition for the outgoing graviton): 
\begin{align}\label{disconnectedCut}
    \MKMOC_{\text{disc}} &= - i \kappa^2\mathcal{M}^{\text{tree}}
\left[
m_2^2 w_2^2 I(u_3, k) + 
m_1^2 w_1^2 I(u_4, k)
\right]\,,
\end{align}
where we used that in the classical limit $-p_{3,4}\cdot k =-p_{2,1}\cdot k + {O}(q^2)\simeq m_{2,1}w_{2,1}$ and $I(u_{3}, k)$ is defined as
\begin{align}
I(u_3, k) &= \frac{1}{m_2}\int \frac{\dd^d \ell}{(2\pi)^d} \frac{\hdelta(2 u_3\cdot \ell) \hdelta(\ell^2)\Theta(\ell^0)}{2\ell\cdot k } \nonumber\\
&= \frac{1}{m_2\omega}  
\int \frac{\dd^d \ell}{(2\pi)^d} \frac{\hdelta(2 u_3\cdot \ell) \hdelta(\ell^2)\Theta(\ell^0)}{2\ell\cdot {\hat k} }\ ,
\label{eq:I3}
\end{align}
where ${\hat k} = k/\omega$, and $I(u_4, k) $ is obtained simply by relabeling.
We did not include $\Theta(p_3^0+\ell^0)$ for the matter cut line because $|\ell^0| \ll p_3^0$ so $\Theta(p_3^0+\ell^0)=1$.
We have also suppressed the $\ie$ in the propagator since it does not affect the integral.
However, to properly include the zero energy graviton contribution across the cut, we need to introduce additional regularizations.\footnote{Integrals of type~\eqref{eq:I3} have appeared in the treatment of rapidity logarithms in quantum field theory, see e.g.~\cite{Chiu:2011qc, Chiu:2012ir}. An additional regularization, effectively introducing a dimensionful regulator which renders the integral nonvanishing, was introduced there as well.}

Since the zero energy graviton is supported only at the origin of phase space, we must regularize it so that this point, $|\ell|=0$, remains in the integration domain.\footnote{This is analogous with the determination of the snail diagrams in {\em quantum} scattering amplitudes, see e.g.~\cite{Bern:2012uf}.} 
We may do this by relaxing the on-shell condition $u_3^2 = -1$ by introducing $u_3\rightarrow \tilde{u}_3 = u_3 - \beta {\hat k}$ with arbitrary $\beta$. 
In the presence of $\hat\delta(\ell^2)$, this modifies the cut 
matter propagator to be 
\begin{align}
    2u_3\cdot\ell = (u_3 + \ell)^2+1 &\rightarrow (\tilde{u}_3+\ell)^2 + 1 \nonumber\\
    &\simeq 2u_3\cdot\ell - 2\beta u_3\cdot\hat{k}\, ,
\end{align}
where we dropped the term proportional to ${\hat k}\cdot \ell$ as it leads to integrals proportional the first of Eq.~\eqref{eq:zero_int} and thus vanishes.
We therefore consider instead the integral
\begin{align}
    \tilde{I}(u_3,k) &= \frac{e^{\euler \epsilon}}{m_2\omega}\int\frac{\dd^d \ell}{(2\pi)^d}\frac{\hdelta(2u_3\cdot\ell - 2\beta u_3\cdot \hat{k})\hdelta(\ell^2)\Theta(\ell^0)}{2\ell\cdot \hat{k}}\nonumber\\
    &=\frac{\Theta(\beta)}{16\pi m_2 w_2}\left[\frac{1}{\epsilon}-\log\frac{w_2^2}{\omega^2} - \log\frac{\beta^2}{\pi}\right],
\end{align}
and we keep only the $O(\epsilon^0)$ terms.
The original integral $I(u_3,k)$ is recovered at $\beta=0$. Using $\Theta(0)=\frac{1}{2}$, we get,\footnote{Alternatively, we can obtain $I(u_3,k)$ directly as a discontinuity of an off-shell triangle integral, which was computed in Eq.~(C.4) of Ref.~\cite{Abreu:2015zaa} (see Ref.~\cite{Bourjaily:2020wvq} for comments on similar integrals), with the same off-shell continuation as discussed here.}
\begin{align}
I(u_3, k) = \frac{1}{32 \pi m_2 w_2}
\left[\frac{1}{\epsilon}-\log \frac{w_2^2}{\omega^2} -\log\frac{\beta^2}{\pi}\right]
\ ,
\label{eq:valueOfCut}
\end{align}
where the divergent constants $\log\beta$ and $1/\epsilon$ will be absorbed by a time shift. 
We note that both integrals in \cref{eq:zero_int} remain zero even if we used the above off-shell regularization.

Putting together \cref{disconnectedCut,eq:valueOfCut}, we find that the cut in \cref{disconnected5ptCut} gives a finite contribution $\mathcal{S}^{\text{1 loop disc}}$, 
\begin{align}
& \MKMOC_{\text{disc}} =\mathcal{S}^{\text{1 loop disc}}-i\omega GE \left[\frac{1}{\epsilon}-\log\frac{\beta^2}{\pi}\right]\mathcal{M}^{\text{tree}}\,,\nonumber\\
& \mathcal{S}^{\text{1 loop disc}} = i G \left[
m_1 w_1 \log\frac{w_1^2}{\omega^2} + m_2 w_2 \log\frac{w_2^2}{\omega^2}
\right] \mathcal{M}^{\text{tree}}\,.
\label{disconnectedCutfinal}
\end{align}
The second term of $\MKMOC_{\text{disc}}$ can be removed by a time shift that supplements the one in \cref{timeshift_IR}.
The first term, $\mathcal{S}^{\text{1 loop disc}}$, which may be interpreted as a direction-dependent time shift, is in fact the BMS supertranslation identified in~\cite{Veneziano:2022zwh} as relating the ``canonical'' and ``intrinsic'' BMS frames, thus explaining the angular momentum loss at ${O}(G^2)$~\cite{Damour:2020tta}, as well as the transformation needed to relate the first nonuniversal term in the soft expansion of the amplitude-based~\cite{Georgoudis:2023eke, QCDmeetsGravityTalk} and MPM waveforms~\cite{Bini:2023fiz}.

We therefore see that zero-energy gravitons generate on the one hand a time-independent background metric at ${O}(G^0)$ (see Sec.~\ref{sec:constant_background}) and, on the other hand, a time-dependent waveform at higher orders in $G$. Moreover both of them 
are needed (see below) to get direct agreement with the MPM waveform, and both of them can simultaneously be removed by the same BMS supertranslation that connects the intrinsic and canonical frames. 
This reinforces the close connection between zero-energy gravitons and the choice of BMS frame, which echoes features in the angular momentum loss  calculations~\cite{Veneziano:2022zwh, Manohar:2022dea, DiVecchia:2022owy}, and suggests that one may consistently ignore the disconnected cut contributions and restore them through the 
BMS transformation of Ref.~\cite{Veneziano:2022zwh}.
It would be interesting to attempt to formulate the observable-based formalism in the ``intrinsic'' BMS frame in terms of a vacuum dressed with a coherent state of zero-energy gravitons in the spirit of Ref.~\cite{Cristofoli:2022phh}. Such a formulation, in which the contribution of zero-energy particles (and the associated subtleties) will emerge because of this dressing, may give a means to argue for an all-order connection between the disconnected cut contributions and the choice of BMS frame.

\section{PN and soft expansions of KMOC waveform}

The one-loop amplitude part of the $\MKMOC$ matrix element computed in \cite{Herderschee:2023fxh, Georgoudis:2023lgf, Brandhuber:2023hhy} exhibits spurious poles (SP). Some of these poles cancel manifestly when the amplitude is expanded in special kinematic limits such as at low velocity, in the soft limit, etc.
However, manifesting such cancellations for general kinematic configurations is, at first sight, highly non-trivial and appears to require the use of Gram determinant and Bianchi identities among a huge number of terms.
Algorithms for addressing such problems have been developed in Ref.~\cite{Abreu:2019odu} (see also Ref.~\cite{Heller:2021qkz}) and applied to the current problem in Ref.~\cite{Bohnenblust:2023qmy}.

In \cref{app:reorg} we take a different route and reorganize the one-loop five-point scattering amplitude~\eqref{eq:amp1L} and the cut term to make manifest the cancellation of the spurious poles on the physical sheet.
By direct inspection of Feynman integrals involving non-trivial numerators, we find combinations of the square roots and logarithms appearing in~\eqref{eq:amp1L} such that they have high-order zeros in correspondence with the singularities of the rational functions multiplying them. 
It turns out that the same functions are sufficient to render both the amplitude and the cut contributions to $\MKMOC$ free of spurious poles.
Their expressions after this reorganization are
\begingroup
\allowdisplaybreaks
\begin{align}
    \label{eq:amp_org}
    \mathcal{M}_{1}^{(1)\text{ fin}} &= \left[ \frac{iw_1}{32\pi}\log\frac{w_1^2}{\muIR^2} + \frac{w_1}{32} + \frac{\Gamma w_1}{64}\right]\mathcal{M}^{(0)}_{d=4} \\*
    &\quad + \frac{iw_1}{32\pi}\mathcal{M}^{(0)}_{\text{extra}} + c_1 I_1 + c_2 I_2 + l^\prime_q L_q + l^\prime_w L_w \nonumber\\*
    &\quad + R + l_q \log\frac{q_1^2}{q_2^2} + l_w \log\frac{w_2^2}{w_1^2} + l_y \frac{\arccoshy}{\sqrt{\y^2-1}} \ ,\nonumber\\
    \label{eq:cut_org}
    \mathcal{S}_{1}^{(1)\text{ fin}} &= \frac{i w_1 \Gamma}{64\pi} \mathcal{M}^{(0)}_{d=4} \log\frac{w_1 w_2 \muIR^2}{q_1^2 q_2^2 (\y^2-1)}  \\*
    &\quad - \frac{i w_1 \Gamma}{64\pi}\mathcal{M}^{(0)}_{\text{extra}} + c_{\text{rat}} + c_{\text{ratio}} \log\frac{w_1^2 w_2^2}{q_1^2 q_2^2} \nonumber\\*
    &\quad  + c_{\y}\arccoshy + c_{q} L_q  + \mathcal{S}_{1}^{(1)\text{ local}}\,,\nonumber
\end{align}
\endgroup
where $\{I_1,I_2,L_q,L_w\}$ are SP-free combinations. Their explicit forms and construction are given in \cref{app:reorg}.
Here $\mathcal{S}_{1}^{(1)\text{ local}}$ includes all terms which have no singularities in $q_{1,2}^2$ (including $c_w L_w$) and therefore integrate to delta-function-supported terms in impact parameter space. Thus $\mathcal{S}_{1}^{(1)\text{ local}}$ can be ignored. The explicit expressions of the $c$ and $l$ coefficients in Eqs.~\eqref{eq:amp_org} and \eqref{eq:cut_org} are included in the ancillary file \texttt{WaveformReorganisation.nb}.

\subsection{Impact-parameter-space waveform}
\label{sec:impact_parameter_waveform}

With the reorganized parts of the momentum space $\MKMOC$ matrix element, we now discuss the Fourier transform~\eqref{eq:MtoWEFT} to obtain the impact-parameter-space (and frequency-space) waveform $\WEFT (k,b_1,b_2) $. Here we will focus on the $\omega \neq0$   (time-dependent) contribution up to $O(G^2)$.
As $\MKMOC(k,q_1,q_2)$ is a function of the kinematic data of the incoming state, the waveform $\WEFT(k,b_1,b_2)$ is naturally given in the \incom frame $(e_0,e_1,e_2,e_3)$ anchored on the incoming momenta $p_1$ and $p_2$.
The unit time vector $e_0$ can be chosen as
\begin{align}
    e_0^{\mu} = \frac{p_1^{\mu}+p_2^{\mu}}{|p_1^{\mu}+p_2^{\mu}|}\,,
\end{align}
while the incoming momenta are along the $e_2$ direction,
\begin{align}\label{eq:com_in}
    p_1 = E_1 \, e_0 + P_{\text{c.m.}} \, e_2\,,\quad 
    p_2 = E_2 \, e_0 - P_{\text{c.m.}} \, e_2\,,
\end{align}
where $E_a = \sqrt{P_{\text{c.m.}}^2 + m_a^2}$. Note that we can express $E_a$ and $P_{\text{c.m.}}$ in terms of $y$ as 
\begin{align}
    P_{\text{c.m.}} &= \frac{m_1 m_2\sqrt{\y^2 -1 }}{E}\,, \nonumber\\
    E &= \sqrt{m_1^2 + 2 \y m_1 m_2 + m_2^2}\,.
\end{align}
We then choose the impact parameter $b_1$ and $b_2$ to satisfy $\frac{E_1}{E}b_1^{\mu} + \frac{E_2}{E}b_2^{\mu} = 0$ and be orthogonal to $e_0$.
We define the relative transferred momentum $q^{\mu}$ by
\begin{align}
    q^{\mu} = q_1^{\mu} - \frac{E_1}{E} k^{\mu} = -q_2^{\mu} + \frac{E_2}{E} k^{\mu}\,,
\end{align}
such that the exponential factor in \cref{eq:MtoWEFT} becomes
\begin{align}
    e^{-i(q_1\cdot b_1+q_2\cdot b_2)} = e^{-iq\cdot (b_1-b_2)} = e^{-iq\cdot b_{12}}\,,
\end{align}
and thus $\WEFT$ depends only on $b_{12}$.
The impact parameter $b_{12}$ is orthogonal to both of the incoming momenta $p_1$ and $p_2$. Thus we can choose it to be along the $e_1$ direction,\footnote{We note that $b_{\text{in}}=b+O(G^2)$. See the discussion below \cref{eq:bbar}.}
\begin{align}
    b_{12} = b_{\text{in}} \, e_1\,.
\end{align}
The momentum $k$ and polarization vector $\varepsilon$ for the emitted graviton are given by
\begin{align}
    \mathbf{\tilde{n}}(\theta,\phi) &= \sin\theta \cos\phi \, e_1 +\sin\theta\sin\phi \, e_2 +\cos\theta \,e_3 \,, \nn\\
    k &= \om \left(e_0 + {\bf\tilde n}(\theta,\phi) \right), \nn \\
    \varepsilon &= \frac{1}{\sqrt{2}} \left[\D_\theta {\bf\tilde n}(\theta,\phi)-  \frac{i}{\sin \theta} \D_\phi {\bf\tilde n}(\theta,\phi)\right].
    \label{eq:Amp_pol_vect}
\end{align}
They are formally the same as \cref{frame_exey}, but the polar angles here $(\theta,\phi)$ are defined instead with respect to the spatial frame $(e_1,e_2,e_3)$.
Therefore, the Fourier transform of $\MKMOC(k,q_1,q_2)$ in \cref{eq:MtoWEFT} now becomes
\begin{align}\label{eq:MtoWEFT2}
    & \FT[f]\equiv\int\mu\Big(k,q+\frac{E_1}{E}k,-q+\frac{E_2}{E}k\Big) e^{-iq\cdot b_{12}}f(q)\,,\nonumber\\
    &\WEFT(\omega,\theta,\phi) = \frac{2i}{\kappa}\, \FT\left[\MKMOC\!\left(k,q+\frac{E_1}{E}k,-q+\frac{E_2}{E}k\right)\right]. 
\end{align}
The resulting impact parameter space waveform $\WEFT$ is in principle a Lorentz invariant function consisting of scalar products of $(\varepsilon,k,p_1,p_2,b_{12})$. However, performing this Fourier transform in complete generality is challenging. 
In the following, we will proceed in the \incom frame, such that the frequency-domain waveform is given as a function $\WEFT(\omega,\theta,\phi)$, and consider its PN expansion.

\subsection{PN expansion}
\label{sec:PNexpansion}

For the PN expansion, we will stick to the conventions of reference \cite{Bini:2023fiz}, also already used here in 
sections \ref{sec:IntroMPM} and \ref{sec:MPMwaveform}. It is an expansion in powers of the velocity parameter
\begin{equation}
    p_\infty \equiv \sqrt{ y^2-1} \ .
\end{equation}
Here $p_\infty$ is taken as dimensionless 
(i.e. we can replace $\eta=\frac1c $, used in Secs. \ref{sec:IntroMPM} and \ref{sec:MPMwaveform} by $1$). 

Following Ref.~\cite{Bini:2023fiz}, one rewrites the kinematic variables in terms of the expansion parameter $p_\infty$, and expands the integrand before computing the Fourier transform to impact-parameter space. This gives rise to a rather simple integrand whose integral can be expressed solely in terms of Bessel K functions and exponentials. 

One starts by decomposing $q^\mu$ in terms of the external vectors (a covariant form of this decomposition, using $u_1, u_2, b$ and a fourth vector orthogonal on them, was introduced in Ref.~\cite{Cristofoli:2021vyo})
\begin{equation}
\label{eq:q_decomposition}
    q^\mu = q^0 e_0^\mu  + \Omega\, Q_x e_1^\mu + q_y e_2^\mu + \Omega\, Q_z e_3^\mu\,,
\end{equation}
where $e_{1,2,3}^\mu$ denote the direction of impact parameter, of the \com momentum and of the orthogonal direction to the scattering plane, respectively -- as defined in Section~\ref{sec:IntroMPM}. Moreover, $\Omega$ is the frequency of the emitted radiation rescaled by the expansion parameter:
\begin{equation}
    \omega = \Omega\, p_\infty\, .
\end{equation}
The integration over $q_{0,y}$ gives the following Jacobian:
\begin{equation}
    \int\!\mu\left(k,q_1,q_2\right) e^{-iq\cdot b_{12}}= \frac{\Omega^2}{4 m_1 m_2 p_\infty} \int \frac{\dd Q_x \dd Q_z}{(2\pi)^2} e^{-i u Q_x}\,,
\end{equation}
where $u$ is the dimensionless frequency variable
\begin{equation}
    u \equiv \Omega \, b\, \equiv \frac{\omega \, b}{p_{\infty}},
\end{equation}
and the delta functions fix the components $q^{0}, q_y$:
\begin{equation}
    \begin{split}
        q^0 &= - \frac{\Omega\, p_{\infty}}{E^2} \left(m_1^2-m_2^2-m_1 m_2 p_{\infty } \sin\theta \sin\phi\right)\,,\\
        q_y &= \frac{\Omega}{E^2} \left(m_1 \sqrt{p_{\infty }^2+1}+m_2\right) \left(m_2 \sqrt{p_{\infty }^2+1}+m_1\right) \,.
    \end{split}
\end{equation}
These relations correspond to Eqs. (5.9) in Ref.~\cite{Bini:2023fiz}, except that the latter equations were expressed in the the barred com frame $(\bar{e}_0,e_x,e_y,e_z)$.
  
For completeness, let us give also the expressions of $w_{1,2}$ in the \com frame:
\begin{equation}
    \begin{split}
        w_1 &= \frac{\Omega\, p_{\infty}}{E}\left[m_1+m_2\left( \sqrt{p_{\infty }^2+1}- p_{\infty} \sin\theta \sin\phi\right)\right]\,,\\
        w_2 &= \frac{\Omega\, p_{\infty}}{E}\left[m_2+m_1\left( \sqrt{p_{\infty }^2+1}+ p_{\infty} \sin\theta \sin\phi\right)\right]\,.\\
    \end{split}
\end{equation}
The original form of the amplitude \cite{Herderschee:2023fxh, Georgoudis:2023lgf, Brandhuber:2023hhy} made it challenging to compute its PN expansion because of the presence of spurious poles, though Ref.~\cite{Bini:2023fiz} succeeded to carry out the expansion up to the 2.5PN approximation, and to 
evaluate the Fourier transform, 
thereby obtaining the EFT waveform as a function of $\om$, $\theta$ and $\phi$. The SP-free version of the amplitude presented above significantly simplifies the task of PN-expanding the amplitude. The leading orders of the various terms in Eq.~\eqref{eq:amp_org} in a small-$p_\infty$ expansion are
\begin{align}
    & c_1 I_1 \sim \pinf^0 & & l'_q L_q \sim \pinf^2 & &R \sim \pinf^{-5} \nonumber\\
    & c_2 I_2 \sim \pinf^{-5} & & l'_w L_w \sim \pinf^2 & &\\
    & l_q \log\frac{q_1^2}{q_2^2} \sim \pinf^0\; & & l_w \log\frac{w_2^2}{w_1^2}\sim \pinf^2\; & & l_y\frac{\arccosh y}{\sqrt{y^2-1}} \sim \pinf^{-1}\,.\nonumber
\end{align}
Thus, in this organization three of the eight terms start contributing only at 2PN order. Moreover, the cancellation of terms $p_\infty^{n\le -2}$ is quite straightforward to observe.
This allowed us to reach the 3PN accuracy. Even though we have so far only computed the 3PN-accurate EFT waveform in the equatorial plane, i.e. when taking $\theta = \frac{\pi}{2}$, there is no technical obstacle to derive it for arbitrary values of $\theta$.
Although most integrals in $Q_z$ and $Q_x$ are absolutely convergent, some of the contributions from the cut term must be interpreted as Fourier transforms of tempered distributions.

   As explained in \cite{Bini:2023fiz} the PN expansion of the amplitude yields an integrand containing only integer or half-integer powers of
    \begin{equation}
        D_0 = 1 + Q_x^2 + Q_z^2\,,
    \end{equation}
    with numerators polynomial in $Q_z$ and $Q_x$. The Fourier transforms of all these terms are computed from the general formula
    \begin{equation}
        \int \frac{\dd Q_x \dd Q_z}{(2\pi)^2} e^{-i u Q_x} \frac{Q_x^n}{D_0^m} = \frac{i^n 2^{-m}}{\pi\, \Gamma[m]}\, \partial_u^n\! \left[u^{m-1} K_{m-1}(u)\right]\,,
    \end{equation}
However, the cut terms (new with this work) involve the first power of $\log D_0$ in the numerator. Such terms can be integrated by using the formula
 \begin{align}
        &\int \frac{\dd Q_x \dd Q_z}{(2\pi)^2} e^{-i u Q_x} \frac{Q_x^n\, \log\! D_0}{D_0^m} =\\
        &=-\partial_\rho \left\{\frac{i^n 2^{-m-\rho}}{\pi\, \Gamma[m+\rho]}\, \partial_u^n\! \left[u^{m+\rho-1} K_{m+\rho-1}(u)\right]\right\} \Bigg|_{\rho=0}\,.\nonumber
    \end{align}
    involving a derivative with respect to the order of the Bessel $K_\nu(u)$ (around an integer value of $\nu$). 
    
    The final result can be further reduced using relations among modified Bessel functions of the second kind to linear combinations of $K_0(u)$, $K_1(u)$ and $e^{-u}$ with coefficients that are rational functions of $u$.
The results of this expansion up to 3PN order are presented in the ancillary file \texttt{PNexpandedEFT.nb}.\footnote{For the analysis of the PN expansion of the various terms contributing to the amplitudes, we refer to Section~V of Ref.~\cite{Bini:2023fiz}.}

\subsection{Soft-graviton expansion}
\label{sec:soft_exp}

We also computed the soft-graviton expansion of the waveform in impact parameter space:\footnote{This PM calculation was carried out independently in Ref.~\cite{Georgoudis:2023eke} and both results were presented for the first time at \cite{QCDmeetsGravityTalk, QCDmeetsGravityTalkRusso}. 
The calculation was done to 2.5~PN order in Ref.~\cite{Bini:2023fiz}.}
\begin{align}\label{softexpan}
    \WEFT\simeq\frac{\mathcal{A}}{\omega} + \mathcal{B}\log\omega + \mathcal{C}\,\omega(\log\omega)^2 + \mathcal{D}\,\omega\log\omega \ .
\end{align}
Care must be take when carrying out the soft-graviton expansion at the level of the waveform integrand. As one quickly discovers, a naive expansion leads to uncontrolled infrared divergences at sufficiently high orders in the subsequent Fourier transform in momentum transfer at sufficiently high orders. 

Thus, the soft-graviton expansion is intertwined with the Fourier transform to impact parameter space. The presence of the small parameter $\omega$ indicates that the appropriate framework for the evaluation of the latter is the method of regions.~\footnote{Note that, while analogous in spirit, is a distinct from the method of regions used to isolate the classical contribution of loop integrals, which have already been been evaluated here. }. There are two distinct contributing regions, $q_{1}\rightarrow \omega^0 q_{1}$ and $q_{1}\rightarrow \omega q_{1}$, and each of them has a different small-$\omega$ expansion which must be added to obtain the correct result.
In both regions the resulting integrals can be evaluated in closed form, as done in Ref.~\cite{Cristofoli:2021vyo}. 

The soft-graviton expansion corresponds to the large time expansion of the time-domain waveform. The behavior of the scattering waveform in this regime exhibits universality as emphasized in Refs.~\cite{Sahoo:2021ctw, Saha:2019tub}. In particular, when expressed in terms of the incoming and outgoing momentum, which differ by the impulse $\Delta p$, the functions $\mathcal{A}$, $\mathcal{B}$ and $\mathcal{C}$ are independent of the specifics of the hard scattering process. We extracted these coefficients from the EFT waveform 
and reproduced the predictions of~Ref.~\cite{Sahoo:2021ctw}. 

We have also computed the non-universal $\mathcal{D}$ coefficient and found exact agreement with Eq.~(9.11) of~\cite{Bini:2023fiz}. 
This is the first order in the soft-graviton expansion where the second region mentioned above, in which the Fourier integration variable scales as $q_1\rightarrow\omega q_1$, first contributes. 
It is also the first order at which the disconnected cut~\eqref{disconnectedCutfinal} contributes crucially to the match with~\cite{Bini:2023fiz}.
The necessity of this term for a successful comparison with the MPM waveform was first suggested in~\cite{Georgoudis:2023eke} as part of a BMS transformation. In Secs.~\ref{sec:constant_background} and \ref{sec:BMSgeneral} we understood its connection to the disconnected cuts of the observable-based formalism if zero-energy gravitons are part of the spectrum. 
On the other hand, the $\epsilon/\epsilon$ term in \cref{eq:MCfinite} does not contribute to any of the coefficients in \cref{softexpan} up to a time shift.

\section{Transforming the KMOC waveform to the barred momentum frame}
\label{sec:rotatingEFT}

The MPM waveform $W_{\rm MPM}(\om, \theta, \phi)$  is computed in a spatial  frame anchored on the classical barred momenta~\eqref{eq:pbar}, while the EFT waveform $\WEFT (\om, \theta, \phi)$ is originally obtained in a spatial frame anchored on the incoming momenta. Both spatial frames are of the \com type, with a negligible $O(G^3)$ difference (due to the radiated momentum) between their time-axes, namely
\begin{align}\label{eq:e0toebar0}
    \bar{e}_0^{\mu} = e_0^{\mu} + O(G^3)\,.
\end{align}
At order $G^2$ the only significant difference between the  \barcom frame spanned by $(\bar{e}_0,e_x,e_y,e_z)$ (defined in \cref{sec:IntroMPM}) and the frame $({e}_0,e_1,e_2,e_3)$ is
a spatial rotation of $\chi/2$ mapping the unit vectors $e_1,e_2$ into $e_x,e_y$. [Note that in both frames we use a polarization vector respectively anchored on $e_1,e_2$ or $e_x,e_y$, as per Eqs. \eqref{frame_exey} and \eqref{eq:Amp_pol_vect}].

Consequently, the function of $\theta$ an $\phi$ describing the EFT waveform referred to the barred momenta frame, say
$ \WEFTb(\omega,\theta,\phi)$, is related to the function of $\theta$ an $\phi$ describing the (original) EFT waveform 
$\WEFT(\omega,\theta,\phi)$ referred to the incoming momenta frame by
\begin{align}\label{eq:chiro}
    \WEFTb(\omega,\theta,\phi) &= \WEFT(\omega,\theta,\phi)  \\
    &\quad + \frac{\chi_{\text{1PM}}}{2}\frac{\partial}{\partial\phi}\WEFT^{G^1}(\omega,\theta,\phi) + O(G^3)\,.\nonumber
\end{align}
Here, as indicated, it is sufficient to insert the 1PM [$O(G^1)$] approximation to the scattering angle, namely
\begin{align}
    \chi_{\text{1PM}} = \frac{2 G E}{b}\frac{2\y^2-1}{\y^2-1}\,.
\end{align}
Thus, using Eq.~\eqref{eq:chiro}, 
the comparison between the MPM and EFT waveform is reduced to comparing the function 
$W_{\rm MPM}(\om, \theta, \phi)$ to the function $\WEFTb(\omega,\theta,\phi)$.

\section{Comparison between MPM and KMOC waveforms}
\label{sec:Comparison}

When working in the barred momenta frame, the comparison between the MPM waveform and the EFT one amounts to comparing the following two functions of  $\om, \theta, \phi$ (and of the impact parameter $b$): 
$W^{\rm MPM}(\om,\theta,\phi)$ on the one hand, and $\WEFTb (\om,\theta,\phi)$ on the other hand. A complete physical agreement between these two waveforms would mean that these two functions differ at most by the effect of an angular-independent time shift $\delta t$. As discussed above, when adding to the EFT waveform the (linearized gravity) contribution coming from the disconnected, zero-frequency diagrams displayed on the left of Fig. \ref{fig:disconnected}, the $O(G^0)$ static background (Coulomb-like) contributions in the two waveforms agree with each other [compare \cref{Win} with \cref{WEFTin}]. In the following, we therefore focus on the frequency-dependent parts ($\om \neq 0$) of the two waveforms, which both start at order $G^1$.

Summarizing the results so far, the function  $\WEFTb (\om,\theta,\phi)$ can be written (at the one loop accuracy)
as the sum of six contributions, namely
\begin{align}\label{eq:chiro2}
    \WEFTb(\omega,\theta,\phi) &= \WEFT^{G^1}(\omega,\theta,\phi) + 
    \WEFT^{\text{1 loop (0)}}(\omega,\theta,\phi) \nn \\
   & +   \WEFT^{\text{1 loop } \, \eps/\eps}(\omega,\theta,\phi)+\WEFT^{\text{1 loop cut}}(\omega,\theta,\phi) 
   \nn \\
   &+ \WEFT^{\text{1 loop disc}}(\omega,\theta,\phi) 
     + \delta^{\text{rot}} \WEFT(\omega,\theta,\phi)  \nn \\
    &+ O(G^3)\,.
\end{align}
Here, each term is defined as follows. 

The $O(G)$ waveform $\WEFT^{G^1}$ comes from the tree-level matrix element $\mathcal{M}^{\text{tree}}$ in \cref{eq:5ptTreeFull},
\begin{align}
    \WEFT^{G^1}(\omega,\theta,\phi) = \frac{2i}{\kappa}\FT\Big[\mathcal{M}^{\text{tree}}\Big].
\end{align}

The $O(G^2)$ (one-loop) contributions to the EFT waveform indicated in Eq. \eqref{eq:chiro2} denote, respectively:

\noindent (i) the contribution from the (linear in $T$) 5-point amplitude obtained by dropping the  IR  $\frac1{\eps}$ divergent
terms in the results of  Refs.~\cite{Brandhuber:2023hhy,Herderschee:2023fxh,Georgoudis:2023lgf}, as  displayed
in Eq. (5.6) of \cite{Bini:2023fiz}, namely, in the notation introduced in Eq.~\eqref{eq:chiro2} above: 
\be
 \WEFT^{\text{1 loop} (0)} = \frac{2i}{\kappa}\,\FT\Big[\kappa^5 m_1^3 m_2^2  \widetilde{\mathcal{M}}_{1}^{(1)}+ (1 \leftrightarrow 2)\Big]\,;
\ee
 (ii) the extra, $\frac{\eps}{\eps}$ contribution from the (linear in $T$) 5-point amplitude generated after absorbing the
 IR  $\frac1{\eps}$ divergence into a time shift as discussed in Secs.~\ref{sec:IR} and \ref{ampANDconnectedcuts}, namely, in the notation introduced in Eq.~\eqref{eq:chiro2} above:
\bea \label{Weps/eps}
 \WEFT^{\text{1 loop } \, \eps/\eps} &=& \frac{2i}{\kappa}\,\FT\Big[\kappa^5 m_1^3 m_2^2 \frac{i w_1}{32\pi}\mathcal{M}^{(0)}_{\text{extra}}+ (1 \leftrightarrow 2)\Big]\nn \\
 &=& \frac{-2 \kappa^4}{32\pi}\, m_1^2 m_2^2  E \om \FT\Big[ \mathcal{M}^{(0)}_{\text{extra}} \Big]\,;
\eea
(iii) the one-loop cut contribution from connected $T$ elements discussed in Sec.~\ref{ampANDconnectedcuts}, namely,  in the notation introduced in Eq.~\eqref{eq:chiro2} above:
\be
\WEFT^{\text{1 loop cut}} = \frac{2i}{\kappa}\,\FT\Big[\mathcal{S}^{\text{1 loop fin}}\Big]\,;
\ee
(iv) the one-loop cut contribution involving disconnected $T$ elements discussed in Sec.~\ref{sec:BMSgeneral},  namely,  in the notation introduced in Eq.~\eqref{eq:chiro2} above:
\be
  \WEFT^{\text{1 loop disc}} = \frac{2i}{\kappa}\,\FT\Big[\mathcal{S}^{\text{1 loop disc}}\Big]\,;
  \ee
  and finally, (v) the contribution from the spatial rotation between the barred-$p$ frame and the incoming $p$ one discussed in Sec.~\ref{sec:rotatingEFT}, namely,  in the notation introduced in Eq.~\eqref{eq:chiro2} above:
\be
 \delta^{\text{rot}} \WEFT=+ \frac{\chi_{\text{1PM}}}{2}\frac{\partial}{\partial\phi}\WEFT^{G^1}(\omega,\theta,\phi)\,.
\ee
Ref.~\cite{Bini:2023fiz} computed, at the $G^2 \eta^5$ accuracy, the (even in $\phi$ sector of the) difference 
\be \label{delWold}
 \delta^{\rm old} W(\omega,\theta,\phi) \equiv W^{\rm EFT'}(\omega,\theta,\phi)-  W^{\rm MPM}(\omega,\theta,\phi),
\ee
between the EFT waveform, as it was understood in the first versions of Refs.~\cite{Brandhuber:2023hhy,Herderschee:2023fxh,Georgoudis:2023lgf}, and interpreted in the barred $p$ frame,
and the MPM one. In the notation of the present work, the quantity  $W^{\rm EFT'}(\omega,\theta,\phi)$ considered in~\cite{Bini:2023fiz} is equal to
\be
W^{\rm EFT'}(\omega,\theta,\phi)=   \WEFT^{G^1}(\omega,\theta,\phi)+\WEFT^{\text{1 loop} (0)}(\omega,\theta,\phi) \,.
\ee
The result of \cite{Bini:2023fiz} for the difference $ \delta^{\rm old} W(\omega,\theta,\phi)$, Eq.~\eqref{delWold}, was
that it started at the $G^2/c^5$ (2.5PN) level, and contained (at this order) both terms linear in the symmetric mass ratio, $\nu$,
and terms quadratic in $\nu$. The explicit expression for the difference
\be
\delta^{\rm old} W(\omega,\theta,\phi)= \delta^\nu W(\omega,\theta,\phi)+ \delta^{\nu^2} W(\omega,\theta,\phi)
\ee
was provided in~\cite{Bini:2023fiz} in tensorial form, and in the time domain. See Eqs. (8.10)-(8.13) there. The
equivalent result in SWSH form, and in the frequency domain, contains only $l=2^+$ contributions at order $\nu^1$,
and both $l=2^+$, $l=3^-$ and $l=4^+$ contributions at order $\nu^2$. Namely,
\bea
 \delta^\nu W(\omega,\theta,\phi) &=& \sum_{m=2,0,-2}  \delta^\nu U_{2m}\; {}_{\bar 2}Y_{lm}(\theta,\phi)\,, \\
 \delta^{\nu^2} W(\omega,\theta,\phi)&=& \sum_{m=2,0,-2}  \delta^{\nu^2} U_{2m}\; {}_{\bar 2}Y_{2m}(\theta,\phi)\nn\\
&+&  \sum_{m=1,-1}  \delta^{\nu^2} V_{3m}\; {}_{\bar 2}Y_{3m}(\theta,\phi) \nn \\
&+& \sum_{m=4,2,0,-2,-4}  \delta^{\nu^2} U_{4m} \; {}_{\bar 2}Y_{4m}(\theta,\phi)\,, \nn \\
\eea
with
\begingroup
\allowdisplaybreaks
\begin{align} \label{delnu1}
\delta^\nu U_{22}&=-\frac{4 i\sqrt{5\pi}}{5}\frac{G^2M^3}{b}   p_\infty^2 \nu u[K_0(u)-2 K_1(u) ] \,,   \nonumber\\*
\delta^\nu U_{20}&= - \frac{4 i\sqrt{30\pi}}{15 }\frac{G^2M^3}{b}   p_\infty^2 \nu u  [K_0(u)+2 uK_1(u)]\,,\nonumber\\*
\delta^\nu U_{2\bar 2}&=-\frac{4 i\sqrt{5\pi}}{5}\frac{G^2M^3}{b}   p_\infty^2 \nu u [K_0(u)+2K_1(u)]\,,\\
\label{delnu2U2}
\delta^{\nu^2} U_{22}&=\frac{8 i\sqrt{5\pi}}{105 }\frac{G^2M^3}{b}   p_\infty^2 \nu^2 u [(13u+6)K_0(u)\nonumber\\*
&\quad + 13(u+1)K_1(u) ]\,,    \nonumber\\*
\delta^{\nu^2} U_{20}&= \frac{8 i\sqrt{30\pi}}{105 }\frac{G^2M^3}{b}   p_\infty^2 \nu^2 u  [2 K_0(u)- u  K_1(u)]\,,\nonumber\\*
\delta^{\nu^2} U_{2\bar 2}&=\frac{8 i\sqrt{5\pi}}{105 }\frac{G^2M^3}{b}   p_\infty^2 \nu^2 u [ (-13u+6)K_0(u)\nonumber\\*
&\quad + 13 (u-1)K_1(u)]\,, \\
\label{delnu2V3}
\delta^{\nu^2} V_{32}&=-\frac{2 i\sqrt{7\pi}}{21}\frac{G^2M^3}{b}   p_\infty^2 \nu^2 u [u K_0(u)\nonumber\\*
& \quad + (1+u) K_1(u)] \,, \nonumber\\*
\delta^{\nu^2} V_{30}&= - \frac{2 i\sqrt{210\pi}}{105 }\frac{G^2M^3}{b}   p_\infty^2 \nu^2 u [u K_0(u)+ K_1(u)]\,,\nonumber\\*
\delta^{\nu^2} V_{3\bar 2}&= -\frac{2 i\sqrt{7\pi}}{21}\frac{G^2M^3}{b}   p_\infty^2 \nu^2 u [u K_0(u)\nonumber\\*
&\quad + (1-u) K_1(u)]\,, \\
\label{delnu2U4}
\delta^{\nu^2} U_{44}&= -\frac{ i\sqrt{7\pi}}{21}\frac{G^2M^3}{b}   p_\infty^2 \nu^2 u [(1+2u)K_0(u)\nonumber\\*
&\quad + 2(u+1) K_1(u)],\nonumber\\*
\delta^{\nu^2} U_{42}&= -\frac{ 2i\sqrt{ \pi}}{21}\frac{G^2M^3}{b}   p_\infty^2 \nu^2 u (u+1) [K_0(u) + K_1(u)],\nonumber\\*
\delta^{\nu^2} U_{40}&= -\frac{ i\sqrt{10\pi}}{105}\frac{G^2M^3}{b}   p_\infty^2 \nu^2 u [3K_0(u)+2u K_1(u)], \nonumber\\*
\delta^{\nu^2} U_{4\bar 2}&= -\frac{ 2i\sqrt{ \pi}}{21}\frac{G^2M^3}{b}   p_\infty^2 \nu^2 u (1-u) [K_0(u)-K_1(u)], \nonumber\\*
\delta^{\nu^2} U_{4\bar 4}&=  -\frac{ i\sqrt{7\pi}}{21}\frac{G^2M^3}{b}   p_\infty^2 \nu^2 u [(1-2u)K_0(u)\nonumber\\*
& \quad + 2(u-1) K_1(u)]\,.
\end{align}
\endgroup

Note that, $W^{\rm EFT'}$ did not (on purpose\footnote{Indeed, one of the important findings of \cite{Bini:2023fiz} was the need to rotate
the EFT to the barred $p$ frame, otherwise the MPM/EFT mismatch would start at the Newtonian order, instead of being
relegated to the 2.5PN order.})
 contain the $\frac{\chi}{2}$ rotation contribution, $ \delta^{\text{rot}} \WEFT$,
nor the three new contributions discussed in the present work, namely
$ \WEFT^{\text{1 loop } \, \eps/\eps}, \WEFT^{\text{1 loop cut}} $ and $  \WEFT^{\text{1 loop disc}} $.
Therefore,  updating the comparison done in \cite{Bini:2023fiz} leads to a new difference
\be 
\delta^{\rm new} W(\omega,\theta,\phi) \equiv W^{\rm EFT}(\omega,\theta,\phi)-  W^{\rm MPM}(\omega,\theta,\phi),
\ee
given by
\bea \label{delWnew}
 \delta^{\rm new} W&=& \delta^{\rm old} W+ \WEFT^{\text{1 loop } \, \eps/\eps}+ \WEFT^{\text{1 loop cut}} \nn \\
& & + \; \WEFT^{\text{1 loop disc}} + \delta^{\text{rot}} \WEFT\,.
\eea
As we shall discuss next, the additional contributions (new with this work) displayed in Eq. \eqref{delWnew} are such
that they precisely cancel the two types ($O(\nu^1)$ and $O(\nu^2)$) of discrepancies that were present in 
Eq. \eqref{delWold}, modulo an angular-independent time-shift between the two waveforms.

\section{KMOC cut and frame rotation}
\label{sec:cut_and_rotation}

It was first noticed in \cite{Bini:2023fiz} by directly comparing the MPM waveform with the amplitude contribution to the KMOC waveform that, up to 2PN order, there is a remarkable agreement between the two when the latter is evaluated in the frame rotated by half the scattering angle relative to the incoming frame (i.e. in the barred frame). As mentioned earlier, this raised the possibility that the cut contribution could be absorbed into this frame change, which is further confirmed in the soft graviton expansion discussed in Sec.~\ref{sec:soft_exp}.

By Fourier-transforming the one-loop {\em connected} cut contribution to impact-parameter space, Ref.~\cite{Georgoudis:2023eke} showed that $\WEFT^{\text{1 loop cut}}$ can be written as a certain differential operator acting on the tree-level waveform up to a further finite redefinition of the the retarded time. In the coordinates used here this differential operator acts as a rotation by $\chi/2$. Using the explicit expression of the connected 
cut~\cite{Herderschee:2023fxh, Brandhuber:2023hhy, Georgoudis:2023ozp}, we explicitly verified this relation,
\begin{align}
    \WEFT^{\text{1 loop cut}} = -\frac{\chi_{\text{1PM}}}{2}\frac{\partial}{\partial\phi} \WEFT^{\text{tree}} + i \omega \delta t^{\text{cut}}  \WEFT^{\text{tree}} \,,
\end{align}
up to 3PN order. The $\epsilon/\epsilon$ term in the second Eq.~\eqref{eq:MCfinite} plays a crucial role.\footnote{As our results were written up, we became aware that this point was also explicitly noted in the version~2 update of Ref.~\cite{Georgoudis:2023eke}.}  
Resumming the 3PN-accurate expansion of the time shift (which starts at Newtonian order) yields the following all-order-in-$\pinf$ expression
\begin{align}
    \delta t^{\text{cut}} = \frac{\kappa^2 E}{32\pi}&\left[\Gamma\log\frac{\muIR b e^{\euler}}{2}-\frac{\y}{2(\y^2-1)^{3/2}}\right.\nonumber\\
    &\;\left.+\;\frac{E_1E_2(2\y^2-1)}{m_1m_2(\y^2-1)^{3/2}}\right] .
\end{align}
To compare with Eq.~(3.48) of the version~1 of Ref.~\cite{Georgoudis:2023eke}, we include the IR-divergent term in \cref{eq:cut1L} to the left hand side (to obtain $\FT[\mathcal{S}^{\text{1 loop}}]$) and the time shift $\delta t_{\text{IR}}^{\text{cut}} = -\frac{GE\Gamma}{2\epsilon}$ that cancels it to the right hand side. This gives,
\begin{align}\label{eq:rotcut}
    \FT\Big[\mathcal{S}^{\text{1 loop}}\Big] &= -\frac{\chi_{\text{1PM}}}{2}\frac{\partial}{\partial\phi}\,\FT\Big[\mathcal{M}^{\text{tree}}\Big] \nonumber\\
    &\quad + i\omega\delta T^{\text{cut}} \,\FT\Big[\mathcal{M}^{\text{tree}}\Big],
\end{align}
where $\delta T^{\text{cut}}=\delta t^{\text{cut}}+\delta t_{\text{IR}}^{\text{cut}}$. Through $O(\epsilon^0)$ our time shift agrees with the result of the version~1 of Ref.~\cite{Georgoudis:2023eke}
\begin{align}\label{GHRform}
    \delta T^{\text{cut}} = \frac{\kappa^2 E}{32\pi} & \left[\frac{\y(\y^2-\frac{3-4\epsilon}{2-2\epsilon})}{(\y^2-1)^{3/2}}\frac{e^{\epsilon\euler}\Gamma(-\epsilon)}{(\muIR^2b^2/4)^{-\epsilon}}\right.\nonumber\\
    & \; \left.+\; \frac{E_1E_2(2y^2-1)}{m_1m_2(\y^2-1)^{3/2}}\right] \,.
\end{align}
Together with the observation that in our \com frame the operator $\bar{\delta}$ in~\cite{Georgoudis:2023eke} becomes $-(\chi_{\text{1PM}}/2)\partial_{\phi}$, we find exact agreement with the prediction of Ref.~\cite{Georgoudis:2023eke} (up to a redefintion of the arbitrary scale $\muIR$), which we have thus verified through direct evaluation up to ${O}(\pinf^3)$, or 3PN order.

In light of this agreement, let us comment on the technical differences between the setup discussed here and that of Ref.~\cite{Georgoudis:2023eke}.
Certain steps in our evaluation and simplification of the cut term, such as removal of terms proportional to the Gram determinant of five vectors, assumed that the external kinematics, including the momentum transfer, were four-dimensional. Consequently, all our Fourier transforms were strictly in four dimensions, as enforced by the decomposition of the momentum transfer in Eq.~\eqref{eq:q_decomposition}. In contrast, the setup in Ref.~\cite{Georgoudis:2023eke} assumes that the Fourier transform is $d$-dimensional. Interestingly, this does not appear to have any consequences in the comparison, such as a difference in the $\epsilon/\epsilon$ contributions that build on the IR-divergent terms. 

The analysis outlined in this section indicates that when comparing the EFT and MPM result in the \barcom frame, we can effectively ignore the 
connected cut contribution $\mathcal{S}^{\text{1 loop}}$ altogether, as it is cancelled by the frame rotation up to a time shift. In other words, 
we can rewrite Eq.~\eqref{delWnew} in the simplified form
\bea 
\label{delWnewnew}
 \delta^{\rm new} W&=& \delta^{\rm old} W+ \WEFT^{\text{1 loop } \, \eps/\eps}+  \nn \\
& +&   \WEFT^{\text{1 loop disc}} + {\rm time\, shift} \ .
\eea
We now proceed to discuss the remaining two terms, $\WEFT^{\text{1 loop } \, \eps/\eps}$ and $ \WEFT^{\text{1 loop disc}}$.

\section{\texorpdfstring{$\WEFT^{\text{1 loop } \, \eps/\eps}$}{W1Loop eps/eps}  and  \texorpdfstring{$\delta^\nu W $}{deltaNuW}}
\label{sec:epBYepANDnu}

The additional $\eps/\eps$ contribution is given by Eq. \eqref{Weps/eps}. Remembering the presence of a factor $1/(m_1 m_2)$ in the Jacobian of $\FT$, $\WEFT^{\text{1 loop } \, \eps/\eps}$ is seen to be proportional to $m_1 m_2$, i.e. to be proportional to $\nu^1$. In addition, as it contains a $O(\eta^3)$ factor $G E \om= \frac{G {\sf M} \om}{c^3}$, and as $\mathcal{M}^{(0)}_{\text{extra}}$ is easily seen to start to differ from its full-gravity tree-level counterpart $\mathcal{M}^{(0)}_{d=4}$ only at the 1PN, 
$O(\eta^2)$, level, we conclude that 
$\WEFT^{\text{1 loop } \, \eps/\eps}$ effectively (i.e. modulo a time shift)
starts contributing to $W$ at the
$G^2 \eta^5$ level. 

An easy computation of  the impact-parameter transform of the PN expansion of $\mathcal{M}^{(0)}_{\text{extra}}$ actually shows that, modulo a time-shift contribution, $i \om \delta t W^{G^1}$, the even-in-$\phi$ projection of
$\WEFT^{\text{1 loop } \, \eps/\eps}$ precisely cancels the $O(\nu)$,
$G^2 \eta^5$-level, contribution
$\delta^\nu W(\omega,\theta,\phi)$ found 
(in \cite{Bini:2023fiz}) to be present in $\delta W(\omega,\theta,\phi)$, and displayed in Eq. \eqref{delnu1}. Namely,
\be \label{nu1comp}
\delta^\nu W + \left[\WEFT^{\text{1 loop } \, \eps/\eps} + i \,\om \,\delta t^{\eps/\eps} \,W^{G^1}\right]^{\phi-{\rm even}}= O\left(\frac{G^2}{c^6}\right)\,,
\ee
with $\delta t^{\eps/\eps}=\frac{G\,E}{c^5}=\frac{GM}{c^3} (1+ \frac{\nu}{2} \pinf^2 +O(\pinf^4))$.

\section{ \texorpdfstring{$\WEFT^{\text{1 loop disc}}$}{W1LoopDisc}  and  \texorpdfstring{$\delta^{\nu^2} W$}{deltaNu2W}}
\label{sec:discANDnusq}

The final terms to discuss (at the highest
PN accuracy explicitly studied in \cite{Bini:2023fiz}) are the (even-in-$\phi$ projection of the) disconnected cut contribution to $\WEFT$
and the $O(\nu^2)$,
$G^2 \eta^5$-level, contribution
$\delta^{\nu^2} W(\omega,\theta,\phi)$ 
displayed in Eqs.~\eqref{delnu2U2}, \eqref{delnu2V3} and \eqref{delnu2U4}.

It was suggested in Ref.~\cite{Georgoudis:2023eke} that the difference $\delta^{\nu^2} W(\omega,\theta,\phi)$ is genuinely present in the EFT-MPM difference and corresponds to the fact that the EFT waveform does not contain some of the zero-frequency-graviton effects contained in the MPM waveform, because it works in the canonical BMS frame of \cite{Veneziano:2022zwh} (in which $W(t=-\infty)=0$) rather than in the intrinsic BMS frame selected both by 
standard classical PM theory and the MPM formalism. However, we found in Sec.~\ref{sec:waveform_KMOC} that disconnected contributions to the cut term account for this. They lead to the additional term $\WEFT^{\text{1 loop disc}}$ in the EFT waveform whose explicit form is 
\be \label{dtVVW}
\WEFT^{\text{1 loop disc}} = + i G \left[
m_1 w_1 \log\frac{w_1^2}{\omega^2} + m_2 w_2 \log\frac{w_2^2}{\omega^2}
\right] \WEFT^{G^1}.
\ee
This term is $O(\nu^2)$, and its PN expansion  starts at the
$G^2/c^5$ level. An easy computation shows that its $\phi$-even projection exactly cancels, at this order,
all the $l=2^+, l=3^-$ and $l=4^+$ pieces
of $\delta^{\nu^2} W(\omega,\theta,\phi)$ 
displayed in Eqs.~\eqref{delnu2U2}, \eqref{delnu2V3} and \eqref{delnu2U4}:
\be \label{nu2comp}
\delta^{\nu^2} W + \left[\WEFT^{\text{1 loop disc}}\right]^{\phi-{\rm even}}= O\left(\frac{G^2}{c^6}\right)\,,
\ee
without the need of an additional time-shift.
 
In other words, after factoring out the effect of the $\chi/2$ rotation, the total time-shift needed to align the MPM waveform (defined as in \cite{Bini:2023fiz}) and the EFT one (completed by the $\eps/\eps$  and the disconnected contributions) is $\delta t^{\rm tot}= G \ME/c^3$. This time-shift can be absorbed in a modification of the link between the time scale $b_0$ entering the tail logarithms of the MPM formalism and the frequency scale $\muIR$ of the amplitudes-based formalism. Namely, Eq. (8.8) of \cite{Bini:2023fiz} is modified into
\be \label{logbmu}
\log(2 b_0 \muIR) + \euler=0\,.
\ee
Note the curious feature that the additional term, \cref{dtVVW}, which reads, 
\be
\WEFT^{\text{1 loop disc}}= i \om \delta t^{\rm VV}(\theta,\phi) W^{G^1}, 
\ee
with the angle-dependent time shift
\be \label{dtVV}
\delta t^{\rm VV}(\theta,\phi) = 2 G \left(m_1 \frac{w_1}{\om} \log\frac{w_1}{\om} + m_1 \frac{w_2}{\om} \log\frac{w_2}{\om} \right)\,,
\ee
{\it doubles}, rather than cancels, the contribution to the EFT waveform coming from the last two lines in Eq. (5.22) of Ref. \cite{Bini:2023fiz}. See further discussion of this term below.

 The results \eqref{eq:rotcut}, \eqref{nu1comp} and  \eqref{nu2comp}  allow us to conclude that the inclusion of the new contributions discussed in this work to the EFT waveform resolves all the mismatches with the even-in-$\phi$ projection of the $G^2 \eta^5$-accurate MPM waveform computed in \cite{Bini:2023fiz}. 
 
 To complete the EFT-MPM comparison on the odd-in-$\phi$ sector, we have also compared the odd-in-$\phi$ projection of $W^{\rm MPM}$,
at the $G^2(\eta^1+\eta^4)$ accuracy (see Eqs. \eqref{V2Geta0} and \eqref{V2G2eta3}) with its EFT counterpart (given, in the
equatorial plane, in the ancillary file \texttt{PNexpandedEFT.nb}).
We found perfect agreement between the two
(with the same time-shift needed for the $\phi$-even part, i.e. the same link Eq. \eqref{logbmu}).
Though our checks in the odd-in-$\phi$ sector probe nonlinear contributions to the waveform less deeply than our even-in-$\phi$ checks, they still include the leading-order tail contributions, and are sensitive to the precise relation Eq.~\eqref{logbmu}.
The explicit comparison between the MPM and amplitudes-based waveforms is presented in the ancillary file \texttt{MPMEFTcomparison.nb}.

\section{Conclusions}
\label{sec:conclusions}

We revisited the amplitude- and observable-based (KMOC~\cite{Kosower:2018adc}) derivation of the classical gravitational wave signal emitted during the scattering of two non-spinning masses at the NNLO order $h_{\mu\nu} =g_{\mu\nu}- \eta_{\mu\nu} \sim G^1+G^2+G^3$, and compared our results
for the frequency-domain, $G$-rescaled waveform quantity
$W(\om, \theta,\phi) \equiv \frac{c^4}{4G} \varepsilon^\mu \varepsilon^\nu h_{\mu\nu} \sim G^0+G^1+G^2$, to the corresponding waveform derived within the (post-Newtonian-matched) Multipolar-post-Minkowskian (MPM) formalism. The recent computation of $W^{\rm MPM}(\om, \theta,\phi)$ \cite{Bini:2023fiz} had pointed out the presence of large discrepancies (starting at the Newtonian level) between $W^{\rm MPM}(\om, \theta,\phi)$ and the (finite part of the) amplitude-based waveform derived in the (first versions of) Refs. \cite{Brandhuber:2023hhy,Herderschee:2023fxh,Georgoudis:2023lgf}, say 
$W^{\rm EFT'}(\om, \theta,\phi)$. Ref. \cite{Bini:2023fiz} had further pointed out that the EFT/MPM mismatch was drastically reduced from the Newtonian level $G^2 \eta^0$ (where $\eta \equiv \frac1c$) to the 2.5 PN level $G^2 \eta^5$ by rotating (by half the classical scattering angle $\chi/2$) the EFT waveform from the incoming momenta frame to the (classical) averaged momenta frame. Part of the puzzle posed by the EFT/MPM mismatch was recently clarified by the eikonal-inspired approach of Georgoudis, Heissenberg  and Russo \cite{Georgoudis:2023eke} who argued that the cut contribution needed \cite{Caron-Huot:2023vxl} in the KMOC framework (and missing in the first derivations of the EFT waveform) was precisely implementing the needed rotation between the incoming-momenta frame (naturally tuned to $W^{\rm EFT}(\om, \theta,\phi)$) and the (classical) averaged-momenta frame (conveniently used to compute $W^{\rm MPM}(\om, \theta,\phi)$) for all final-states that do not contain zero-energy gravitons. Ref.~\cite{Georgoudis:2023eke} compared only the soft expansion of $W^{\rm EFT}(\om, \theta,\phi)$ to the soft expansion of $W^{\rm MPM}(\om, \theta,\phi)$ and could indeed verify that the $\chi/2$ rotational effect of the cut term reduced the EFT/MPM mismatch between these expansions to the  $G^2 \eta^5$ level. 
However, the soft limit is a rather blunt tool which is not sensitive to the details of the EFT/MPM mismatch (as discussed, e.g., in Section IX of \cite{Bini:2023fiz}).

Indeed, after factoring out the effect of the $\chi/2$ rotation, the EFT-MPM mismatch found in \cite{Bini:2023fiz} still contained significant terms of order $\nu^1 G^2 \eta^5$ and of order $\nu^2 G^2 \eta^5$. The terms $O(\nu^1 G^2 \eta^5)$ effectively disappear in the soft limit (because they are equivalent to an unobservable time shift). We showed in the present work that a subtle additional $\eps/\eps$ contribution to the amplitude result (coming, in our scheme, from the 
tracelessness of the physical graviton in $d$ dimensions) precisely cancelled the $O(\nu^1 G^2 \eta^5)$ EFT-MPM mismatch (which could not be seen in the soft limit). 
Concerning the last remaining $O(\nu^2 G^2 \eta^5)$ EFT-MPM mismatch, it was argued by Georgoudis et al. \cite{Georgoudis:2023eke}
that it remained and that, in view of Eqs. \eqref{dtVVW}, \eqref{dtVV}, it was to be physically interpreted as the presence of the particular  $O(G)$ supertranslation (i.e. an angle-dependent time shift)
$\delta t^{\rm VV}(\theta,\phi)$, Eq. \eqref{dtVV}, pointed out by Veneziano and Vilkovisky~\cite{Veneziano:2022zwh}, for its role in transforming
 the ``intrinsic'' BMS frame (naturally used both in classical, worldline PM perturbation theory, and in the MPM formalism) into the ``canonical'' BMS frame (defined so that the initial value of the waveform vanishes in the infinite past). The interpretation of the necessary presence of the supertranslation \eqref{dtVV} was that this supertranslation
removes the effect of static modes (zero-frequency gravitons) in the waveform, and that the amplitude approach naturally computes a waveform without static modes.

Our new results suggest, however, a different picture. Static modes can be incorporated in the KMOC waveform if one includes disconnected $T$-matrix elements,
both at the linear-in-$T$ level and at the bilinear-in-$T$ level (cut term). We have explicitly shown (see Eq.~\eqref{WEFTin}) that the inclusion of disconnected diagrams (pictured in Fig.~\ref{disconnected5ptTree}) to the linear-in-$T$ waveform was precisely generating the initial, zero-frequency contribution which is naturally present in the MPM waveform (see Eq.~\eqref{fin}). 
The situation concerning the effect of disconnected diagrams in the bilinear-in-$T$ term is more subtle. In a natural regularization which takes the matter velocity slightly off-shell, $u_3\rightarrow \tilde u_3$, we showed both by direct evaluation and by interpreting it as the discontinuity of a triangle integral that the cut pictured in Fig.~\ref{disconnected5ptCut} yields an additional contribution to $\WEFTb$ that precisely cancels the effect of the supertranslation \eqref{dtVV}, thereby removing the last remaining $O(\nu^2 G^2 \eta^5)$ EFT-MPM mismatch.
We have also shown in \cref{sec:background2} that other disconnected $T$-matrix elements at one-loop
%
%
would not modify our finding. 
It would clearly be interesting to confirm in a different way our conclusion that the inclusion of disconnected diagrams in the KMOC formalism does incorporate all the physical effects linked to the static modes (which are included in the MPM formalism). 

Our new understanding of the zero-frequency contribution opens up several future research directions.
It would be interesting to know whether a higher-order KMOC 
calculation, systematically ignoring disconnected diagrams, would automatically yield a waveform differing from the MPM one {\it only} by taking into account the full (nonlinear) effect of the supertranslation \eqref{dtVV} (which is fully determined by the $O(G)$ linearized-gravity Coulombic background \eqref{fin}, \eqref{Win}). A classical (worldline-based) PM-gravity computation (\`a la \cite{Kovacs:1977uw,Bel:1981be,Mougiakakos:2021ckm})
of the one-loop waveform would be very valuable.  
Conversely, it would also be interesting to explore if the time-dependent and time-independent contributions of zero-energy gravitons can be captured by a nontrivial vacuum containing a coherent state or condensate of such modes, perhaps along the lines of~\cite{DiVecchia:2022owy,Cristofoli:2022phh,Ware:2013zja}.


In this paper we have  shown the agreement between the MPM and EFT waveforms (over the full celestial sphere) at the $(G+G^2)(\eta^0+\eta^2+\eta^3+\eta^5)$  accuracy (2.5PN order) in the $\phi$-even sector, and at the $(G+G^2)(\eta^1 +\eta^4)$ accuracy in the $\phi$-odd sector. [The agreement of the $\phi$-even $(G+G^2)\eta^4$ (2PN) terms was also separately checked in the equatorial plane.]
The details of our results are presented in three ancillary files: (1) the PN expansions of the spin-weighted-spherical-harmonic coefficients $W^{\rm MPM}_{lm}$ of the MPM waveform $W^{\rm MPM}(\theta,\phi)$ are given in the ancillary file \texttt{MPMmultipoles.nb} so as to reach the accuracy $(G+G^2)(\eta^0+\eta^2+\eta^3+\eta^4+\eta^5)$ for even $m$'s ($\phi$-even sector; which has been extended here to include the 2PN terms) and the accuracy $(G+G^2)(\eta^1 +\eta^4)$ for odd $m$'s ($\phi$-odd sector);
(2) 
the PN-expansion of the $\theta= \frac{\pi}{2}$ restriction of the rotated  one-loop-accurate EFT waveform $ \WEFTb(\omega,\theta,\phi)$, Eq.~\eqref{eq:chiro}, is given in the ancillary file \texttt{PNexpandedEFT.nb} at the 3PN accuracy, i.e. 
$(G+G^2)(\eta^0+\eta^1+\eta^2+\eta^3+\eta^4+\eta^5+\eta^6)$;
and (3) the comparison (and agreement), in the equatorial plane, between the two waveforms is presented in the ancillary file \texttt{MPMEFTcomparison.nb}.

Our current organization of the amplitude and cut contributions, in terms of functions free of spurious poles, makes is possible to proceed to higher PN orders at $O(G^3)$ in the EFT side. By contrast, analogous extensions on the MPM side require (when using the MPM techniques used so far) to compute new, challenging, higher-PN contributions. This suggests that borrowing information from the EFT calculation might help to improve the power of the MPM formalism.
Moreover, it would undoubtedly be interesting to extend our comparison to scattering waveforms for spinning bodies, for which the MPM formalism naturally applies~\cite{Blanchet:2006gy, Blanchet:2013haa} and $O(G^3)$ amplitudes-based results are available~\cite{Bohnenblust:2023qmy}.

The $O(G^3)$ waveform discussed here allows for a direct determination of the $O(G^4)$ angular momentum and energy loss which have nontrivial implications for conservative binary scattering at $O(G^5)$, including information on the first second-self-force contribution to this process. 
Carrying out the next-order waveform calculation, which both gives analogous information $O(G^6)$ and probes the agreement between EFT and MPM at higher orders in Newton's constant, poses an interesting and important challenge to the EFT calculation, through the more involved multi-scale integration-by-parts reduction and the evaluation of the ensuing multi-scale master integrals and of the Fourier-transform to impact parameter space. 

One of the important takeaway messages of our new results is that, after having sorted out the subtleties that were hidden in the EFT one-loop waveform, we obtained a remarkable confirmation that the classical limit of an amplitude-based waveform does correctly incorporate the many subtle classical effects that were included in the $O(G^2 \eta^5)$-accurate MPM waveform such as (i) radiation-reaction effects on the worldlines, (ii) high-multipolarity tail effects in the wave-zone, and (iii) cubically nonlinear multipole couplings in the exterior zone.
The fact that the road leading to the present successful EFT/MPM comparison had some bumps, which taught us interesting lessons, is another example of the useful synergy between amplitude-based, and classical perturbation-theory-based, approaches to gravitational physics.

\begin{acknowledgments}
We thank Holmfridur Hannesdottir, Sebastian Mizera, and Cristian Vergu for discussions; RR and FT thank Enrico Herrmann and Michael Ruf for emphasizing the importance of zero-energy modes. We also thank Alessandro Georgoudis, Carlo Heissenberg and Rodolfo Russo for discussions and communications on the submission. 
The present research was partially supported by the 2021 Balzan Prize for Gravitation: Physical and Astrophysical Aspects, awarded to T. Damour. D.B. acknowledges the sponsorship of the Italian Gruppo Nazionale per la Fisica Matematica (GNFM) of the Istituto Nazionale di Alta Matematica (INDAM), as well as the hospitality and the highly stimulating environment of the Institut des Hautes Etudes Scientifiques. 
S.D.A.'s research is supported by the European Research Council, under grant ERC–AdG–88541.
A.H. is supported by the Simons Foundation.
R.R. and F.T. are supported by the U.S. Department of Energy (DOE) under award number DE-SC00019066.

\end{acknowledgments}

\appendix

\section{Notations}
\label{app:notation}

In this appendix, we collect the various notations used in the main text:
\begingroup
\allowdisplaybreaks
\begin{gather*}
\kappa^2 = 32\pi G \\
\eta_{\mu\nu}={\rm diag}(-1,+1,+1,+1)\\
g_{\mu\nu} = \eta_{\mu\nu} + h_{\mu\nu}=\eta_{\mu\nu} + \kappa \,\mathsf{h}_{\mu\nu} \\
w_1 = -u_1\cdot k \qquad w_2 = -u_2\cdot k \\
\y = -u_1\cdot u_2 \qquad \Gamma = \frac{3\y-2\y^3}{(\y^2-1)^{3/2}} \\
F_{\mu\nu} = k_{\mu}\varepsilon_{\nu} - k_{\nu}\varepsilon_{\mu} \\
F_1 = F_{\mu\nu}q_1^{\mu}u_1^{\nu} \\
F_2 = F_{\mu\nu}q_2^{\mu}u_2^{\nu} \\
F_3 = F_{\mu\nu}u_1^{\mu}u_2^{\nu} \\
E = \sqrt{m_1^2 + 2 \y m_1 m_2 + m_2^2} \\
E_1 = \frac{m_1}{E}(\y m_2 + m_1) \qquad E_2 = \frac{m_2}{E}(\y m_1 + m_2) \\
P_{\text{c.m.}} = \frac{m_1 m_2}{E}\sqrt{\y^2-1} \\
p_{\infty} = \sqrt{y^2-1} \\
M= m_1+m_2 \qquad \nu = \frac{m_1 m_2}{(m_1+m_2)^2}\\
{\sf M} = \frac{E}{c^2} \qquad \eta = \frac{1}{c} 
\end{gather*}
\endgroup
\noindent
The speed of light is set to unity, $c=1$, unless it is useful to emphasize the PN order.

\section{One-loop amplitude and KMOC cut}\label{app:amp}

Here we collect the one-loop amplitude and cut contribution to the KMOC matrix element $\MKMOC$ that are known in the literature. We follow the notation used in~\cite{Brandhuber:2023hhy}, which was also used in~\cite{Bini:2023fiz}. The IR finite part of the classical one-loop five-point amplitude is given by
\begin{align}
    \widetilde{\mathcal{M}}_{1}^{(1)} &= \left[ \frac{iw_1}{32\pi}\log\frac{w_1^2}{\muIR^2} + \frac{w_1}{32} + \frac{\Gamma w_1}{64}\right]\mathcal{M}^{(0)}_{d=4} + \frac{\mathfrak{R}}{\pi} + \mathfrak{c}_1\mathcal{I}_1  \nonumber\\
    &\quad  + \mathfrak{c}_2\mathcal{I}_2 + \frac{\mathfrak{l}_q}{\pi}\log\frac{q_1^2}{q_2^2} + \frac{\mathfrak{l}_{w_2}}{\pi}\log\frac{w_2^2}{w_1^2} + \frac{\mathfrak{l}_{\y}}{\pi}\frac{\arccoshy}{\sqrt{\y^2-1}} \,,
\end{align}
where $\muIR$ is the arbitrary scale introduced by the dimensional regularization, and $\mathcal{I}_{1,2}$ are two triangle integrals,
\begin{align}
    \mathcal{I}_{1} &= -\frac{i}{32\sqrt{q_2^2+w_1^2}} + \frac{1}{16\pi\sqrt{q_2^2+w_1^2}}\arcsinhw1q2\,. \nonumber\\
    \mathcal{I}_2 &= -\frac{i}{32\sqrt{q_1^2}}\,.
\end{align}
The gauge invariant coefficients $\mathfrak{R}$, $\mathfrak{c}_{1,2}$ and $\mathfrak{l}_{q,w_2,y}$ are functions of the kinematic data $w_{1,2}$, $q_{1,2}^2$, $F_{1,2,3}$ and $y$. Their explicit expressions can be found in the ancillary files of~\cite{Brandhuber:2023hhy}. 

The IR finite part of the one-loop KMOC cut is
\begin{align}
    \mathcal{S}^{(1)}_1 &= \frac{i w_1 \Gamma}{64\pi} \mathcal{M}^{(0)}_{d=4}  \log\muIR^2
    + \mathfrak{c}_{\text{rat}} + \mathfrak{c}_{q_1}\log q_1^2 \nonumber\\
    &\quad  + \mathfrak{c}_{q_2}\log q_2^2 + \mathfrak{c}_{w_1}\log w_1^2 + \mathfrak{c}_{w_2}\log w_2^2 \nonumber\\
    &\quad + \mathfrak{c}_{\y}^{(1)} \frac{\log(\y^2 -1)}{\sqrt{\y^2-1}} + \mathfrak{c}_{\y}^{(2)}\arccoshy\,, 
\end{align}
where the $\mathfrak{c}$-coefficients can also be found in the ancillary files of~\cite{Brandhuber:2023hhy}.

\section{Reorganization of the one-loop five-point amplitude free of spurious poles}
\label{app:reorg}

By direct inspection of the rational coefficients appearing in \cref{eq:amp1L} and \cref{app:amp} (which can be found in the ancillary files of \cite{Brandhuber:2023hhy,Herderschee:2023fxh,Bohnenblust:2023qmy}), we find four SPs (which correspond to lower dimensional Gram determinants) in the physical region, plus two additional SPs on the physical sheet (but for complex kinematics). 
The former four are, respectively, given by the vanishing loci of 
\begin{align}
    \Delta_1 &\equiv \y - \frac{w_1^2 + w_2^2}{2 w_1 w_2}\,,\nonumber\\
    \Delta_2 &\equiv \y - \frac{(q_1^2 w_1)^2 + (q_2^2 w_2)^2}{2 q_1^2 q_2^2 w_1 w_2}\,,\nonumber\\
    \Delta_3 &\equiv (q_1^2 - q_2^2)^2 - 4 w_1^2 q_1^2 \,,\nonumber\\
    \Delta_4 &\equiv (q_1^2 - q_2^2)^2 - 4 w_2^2 q_2^2 \, ;
\end{align}
the latter two are located, respectively, at the zeroes of\footnote{A note of caution should be made about these additional singularities. 
The sign of the $i\epsilon$ prescription for $w_1$ determines whether or not the integral $\mathcal{I}_1$ has a branch point at $\Pi_1$ in the complex $q_1^2$ plane. We use the prescription $w_1 \to w_1 + i \epsilon$, which does not give a singularity at $\Pi_1$. The SP also cancels among the coefficient of $\mathcal{I}_1$ and the rational terms $\mathfrak{R}$.}
\begin{equation}
    \begin{split}
        \Pi_1 \equiv w_1^2 + q_2^2 \,,\\
        \Pi_2 \equiv w_2^2 + q_1^2 \,.
    \end{split}
\end{equation}

At $\Delta_{1,2}=0$, we find second-order poles in the rational coefficients $\mathfrak{l}_{q,w,\y}$ and simple poles in $\mathfrak{R}$. We first focus on $\Delta_{1}$ and then trivially generalize the result to analogous poles at $\Delta_{2}=0$ through the substitution $w_i \to w_i q_i^2$, which maps $\Delta_1$ into $\Delta_2$. 

The cancellation of the second-order poles requires that we construct combinations of square roots and logarithms that are proportional to $\Delta_1^2$. To understand the pattern, we expand $\arccosh\y$ around $\Delta_{1} =0$:
\begin{align}
    \frac{\arccosh\y}{\sqrt{\y^2-1}} =& \left(\frac{2 z}{z^2-1} -\frac{4 z^2 (1+z^2)}{(z^2-1)^3}\Delta_1\right)\log z \nonumber\\
    &+ \frac{4 z^2 }{(z^2-1)^2}\Delta_1 + O(\Delta_1^2)\,,
\end{align}
where $z=\frac{w_2}{w_1}$ and the expansion is symmetric in $z\to \frac{1}{z}$.
This expansion indicates that the following combination of logarithms,
\begin{widetext}
\begin{equation}
    \log z - \frac{z^2-1}{2z}\left[\left(1+\frac{\y \Delta_1}{\y ^2-1}\right)\frac{\arccosh\y}{\sqrt{\y^2-1}} 
    -\frac{\Delta_1}{\y ^2-1}\right] = O(\Delta_1^2)\, ,
\end{equation}
should lead to expressions that are free of second-order SPs in $\Delta_1$ and $\Delta_2$. This is indeed the case and we thus found 
the first two desired functions:
\begin{align}
    \label{eq:Lw}
    L_w &= \frac{1}{\Delta_1^2}\left\{2\log\frac{w_2}{w_1}-\frac{w_2^2 -w_1^2}{w_1 w_2}\left[\left(1+\frac{\y \Delta_1}{\y ^2-1}\right)\frac{\arccoshy}{\sqrt{\y^2-1}} - \frac{\Delta_1}{\y^2-1}\right]\right\}\,,\\
    \label{eq:Lq}
    L_q &= \frac{1}{\Delta_2^2}\left\{2\log\frac{q_2^2}{q_1^2}+2\log\frac{w_2}{w_1}-\frac{(q_2^2 w_2)^2 -(q_1^2 w_1)^2}{q_1^2 q_2^2 w_1 w_2}\left[\left(1+\frac{\y \Delta_2}{\y ^2-1}\right)\frac{\arccoshy}{\sqrt{\y^2-1}} - \frac{\Delta_2}{\y^2-1}\right]\right\}\,.
\end{align}
We stress that such functions are not unique since any term $O(\Delta_i^0)$ can be added without changing the desired properties. 

We next turn our attention to the poles in $\Delta_3$ and $\Pi_1$, which are closely related and are of fourth and third order, respectively, in the part of the amplitude proportional to $m_1^3 m_2^2$. 
When constructing an ansatz for SP-free functions we should account for the fact that $\Pi_1$ appears only in the coefficient of $\mathcal{I}_1$ and in $\mathfrak{R}$ and thus we should not include it together with $\mathcal{I}_2$ and $\log\frac{q_1^2}{q_2^2}$. Further accounting for the order of the pole in $\Pi_1$, a suitable ansatz is
\begin{equation}
\frac{\Delta_3 P_1}{(q_1^2)^4 w_1^5 \Pi_1^3}+\frac{P_2}{(q_1^2)^4 \Pi_1^3 }\mathcal{I}_1+\frac{P_3}{(q_1^2)^4 w_1^4}\log\frac{q_1^2}{q_2^2}+\mathcal{I}_2\, .
\end{equation}
In this expression $P_i$'s are arbitrary polynomials in $q_1^2$, $q_2^2$ and $w_1$; they are fixed by demanding that this expression has a four-th order zero at $\Delta_3=0$ and cubic zero at $\Pi_1=0$ (which is away from the physical region). Thus, a function with no SPs in $\Delta_3$ is
\label{eq:I1}
\begin{align}
    I_1 &= - \frac{q_1^2 + q_2^2}{256\pi (q_1^2)^2 w_1 \Pi_1\Delta_3}\left(\frac{3(q_2^2)^2 + 11q_2^2 w_1^2 + 23w_1^4}{384(q_1^2 w_1^2)^2\Pi_1^2}-\frac{q_2^2+4w_1^2}{16 q_1^2 w_1^2\Pi_1\Delta_3}+\frac{1}{\Delta_3^2}\right) \nonumber\\
    &\quad + \frac{q_1^2 + q_2^2}{128(q_1^2)^4\Pi_1^3\Delta_3}\left[\frac{35}{16} - \frac{35(q_1^2 + q_2^2)^2}{8\Delta_3} + \frac{7(q_1^2 + q_2^2)^4}{2\Delta_3^2} - \frac{(q_1^2 + q_2^2)^6}{\Delta_3^3}\right]\mathcal{I}_1\nonumber\\
    &\quad - \frac{(q_1^2-q_2^2)\log\frac{q_1^2}{q_2^2}}{64\pi q_1^2 w_1 \Delta_3}\left[\frac{5}{1024(q_1^2 w_1^2)^3} - \frac{3}{128(q_1^2 w_1^2)^2\Delta_3}+\frac{1}{8q_1^2w_1^2\Delta_3^2}-\frac{1}{\Delta_3^3}\right]+\frac{\mathcal{I}_2}{\Delta_3^4}\,.
\end{align}    
\end{widetext}
Finally, a function with no SPs in either $\Delta_3$ or in $\Pi_1$ is
\begin{align}
\label{eq:I2}
    I_2 = \frac{1}{\Pi_1^3}\left(\mathcal{I}_1 - \frac{(q_2^2)^2}{80\pi w_1^5} - \frac{11q_2^2}{240\pi w_1^3} - \frac{23}{240\pi w_1}\right)\, .
\end{align}
$I_1$ and $I_2$ will render SP-free the part of the classical amplitude that is proportional to $m_1^3 m_2^2$. The combined transformations $q_1\leftrightarrow q_2$ and $w_1\leftrightarrow w_2$ yield the functions for the part of the amplitude that is proportional to $m_1^2 m_2^3$.

To expose the functions $L_q$, $L_w$, $I_1$ and $I_2$ in the expression of the amplitude, we use partial fractioning identities,\footnote{For this purpose we use the \texttt{Mathematica} package \texttt{MultivariateApart} \cite{Heller:2021qkz}.} to isolate the SPs in $\mathfrak{c}_{1,2}$, $\mathfrak{l}_{q,w,\y}$ and $\mathfrak{R}$ and further add and subtract appropriate combinations of logarithms and square roots. 
Their coefficients are, by construction, free of SPs. 
We then verified that the coefficients of the remaining logarithms are also free of SPs upon applying the Bianchi and four-dimensional identities.
While all coefficients and rational terms are manifestly free of the initial SPs, this feature comes at the expense of introducing higher powers of the physical poles.

The final expressions are given in Eqs.~\eqref{eq:amp_org} and \eqref{eq:cut_org} in the main text, and explicit expressions of the $c$ and $l$ coefficients there are included in the ancillary file \texttt{WaveformReorganisation.nb}.

\section{Disconnected matrix elements supported by zero energy graviton}
\label{sec:background2}

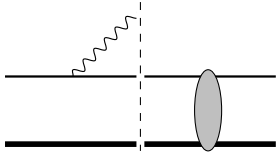
\begin{figure}[t]
    \centering
    \begin{tikzpicture}
        \pgfmathsetmacro{\sc}{0.9};        
        \path (0,0) pic [scale=\sc] {discOneLoopCut};
    \end{tikzpicture}
    \caption{The unitarity cut that only has support on the external zero energy graviton.}    
    \label{fig:oneloopbg}
\end{figure}

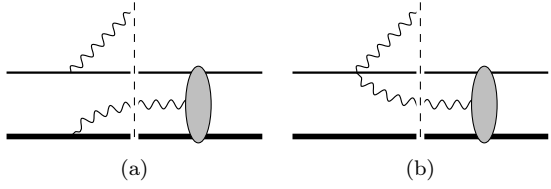
\begin{figure}[t]
    \centering
    \subfloat[]{\label{fig:bg2a}\begin{tikzpicture}
        \pgfmathsetmacro{\sc}{0.85};        
        \path (0,0) pic [scale=\sc] {discOneLoopCutDS2};
    \end{tikzpicture}}\quad
    \subfloat[]{\label{fig:bg2b}\begin{tikzpicture}
        \pgfmathsetmacro{\sc}{0.85};        
        \path (0,0) pic [scale=\sc] {discOneLoopCutDS};
    \end{tikzpicture}}
    \caption{The unitarity cuts that only have support when both the external and one internal graviton have zero energy.}    
    \label{fig:oneloopbg2}
\end{figure}

In this appendix, we will discuss the role of the cut contributions to $\MKMOC$ that have support on the external zero energy graviton, which could potentially modify the constant background. Besides \cref{disconnected5ptTree} that leads to the Schwarzschild background, other possible contributions to the background, given in \cref{fig:oneloopbg,fig:oneloopbg2},
all involve a cut.

We start with the cut in \cref{fig:oneloopbg}. It is more convenient here to use the true momenta of the massive particles. The cut is given by
\begin{align}
    \mathcal{S}_5 = -i\kappa \Big[(\varepsilon\cdot \mathsf{p}_1)^2 \hdelta(2\mathsf{p}_1\cdot k) + (1\leftrightarrow 2)\Big]\mathcal{M}_4\,.
\end{align}
We note that the true momentum $\mathsf{p}_a$ is related to $p_a$ used in the main text through $p_a = \mathsf{p}_a+q_a/2$. Since the unitarity cut forces the external graviton to have vanishing energy, it will interact with the five-point amplitude in the soft limit,
\begin{widetext}
\begin{align}
    \mathcal{M}_5\Big|_{\text{soft}} &= \kappa\left[-\frac{(\varepsilon\cdot \mathsf{p}_1)^2}{2\mathsf{p}_1\cdot k+\ie}-\frac{(\varepsilon\cdot \mathsf{p}_2)^2}{2\mathsf{p}_2\cdot k+\ie}+\frac{(\varepsilon\cdot\mathsf{p}_3)^2}{2\mathsf{p}_3\cdot k-\ie}+\frac{(\varepsilon\cdot\mathsf{p}_4)^2}{2\mathsf{p}_4\cdot k-\ie}\right]\mathcal{M}_4 \\
    &=\kappa\left[i(\varepsilon\cdot\mathsf{p}_1)^2\hdelta(2\mathsf{p}_1\cdot k)-\frac{2(\varepsilon\cdot\mathsf{p}_1)(\varepsilon\cdot q_1)-(\varepsilon\cdot q_1)^2}{2\mathsf{p}_1\cdot k-\ie}+\frac{(\varepsilon\cdot \mathsf{p}_1-\varepsilon\cdot q_1)^2}{2\mathsf{p}_1\cdot k-\ie}\sum_{n=1}^{\infty}\left(\frac{2q_1\cdot k}{2\mathsf{p}_1\cdot k-\ie}\right)^{n} + (1\leftrightarrow 2)\right]\mathcal{M}_4 \,.\nonumber
\end{align}
\end{widetext}
We see that the delta functions cancel in the combination $\mathcal{M}_5 + \mathcal{S}_5$ in the soft limit such that there are no contributions to the constant background. What is left is given in terms of retarded propagators. They will generate instead the memory at $t\rightarrow+\infty$. This statement works for more generic cases as we do not assume a particular loop order for $\mathcal{M}_4$. In principle, the same conclusion can be drawn from using $p_a$ momenta. The retarded matter propagators would then result from a cancellation between (derivatives of) delta functions and principal values, much like Ref.~\cite{Caron-Huot:2023vxl}.

Finally, we sketch the derivation that the two cuts shown in \cref{fig:oneloopbg2} also vanish. For both cuts, we can use the soft limit of the five-point amplitude in the right blob and sew it to the left blob. One will find that \cref{fig:bg2a} only reduces to homogeneous integrands as in \cref{eq:zero_int}, which integrate to zero. For \cref{fig:bg2b}, the only additional nontrivial integral is
\begin{align}
    \int\dd^d \ell \frac{\hdelta(u_1\cdot\ell)\hdelta(u_2\cdot k + u_2\cdot \ell)\delta(\ell^2)\Theta(\ell^0)}{k\cdot\ell}\,.
\end{align}
One can rescale $\ell\rightarrow\omega\ell$ and find that the integral is proportional to $\omega^{d-4}\delta(\omega)$. 
However, because this integral comes from the diagram with a massless three-point vertex, signaled by the propagator $(k\cdot\ell)^{-1}$, the coefficient should scale as $\omega^2$. Therefore, overall we have the scaling $\omega^{d-2}\delta(\omega)$, which vanishes in four dimensions.

\bibliographystyle{utphys.bst}
\bibliography{draft}

\end{document}